\documentclass[12pt,preprint]{aastex}
\usepackage{amsmath}

\def\mbf#1{\mbox{\boldmath ${#1}$}}

\begin{document}


\title{Coupled evolutions of the stellar obliquity, orbital distance, and planet's
radius due to the Ohmic dissipation induced in a diamagnetic hot Jupiter around a
magnetic T Tauri star }


\author{Yu-Ling Chang\altaffilmark{1,2},Peter H. Bodenheimer\altaffilmark{3},and
Pin-Gao Gu\altaffilmark{1}}



\altaffiltext{1}{Institute of Astronomy \& Astrophysics, Academia Sinica,
    Taipei 10617, Taiwan}
\altaffiltext{2}{Graduate Institute of Astronomy, National Central University,
Jhongli 32001, Taiwan} \altaffiltext{3}{UCO/Lick Observatory, University of
California, Santa Cruz, CA 95064, USA}


\begin{abstract}
We revisit the calculation of the Ohmic dissipation in a hot Jupiter presented in
\citet{Laine} by considering more realistic interior structures, stellar obliquity,
and the resulting orbital evolution. In this simplified approach, the young hot
Jupiter of one Jupiter mass is modelled as a diamagnetic sphere with a finite
resistivity, orbiting across tilted stellar magnetic dipole fields in vacuum. Since
the induced Ohmic dissipation occurs mostly near the planet's surface, we find that
the dissipation is unable to significantly expand the young hot Jupiter.
Nevertheless, the planet inside a small co-rotation orbital radius can undergo
orbital decay by the dissipation torque and finally overfill its Roche lobe during
the T Tauri star phase. The stellar obliquity can evolve significantly if the
magnetic dipole is parallel/anti-parallel to the stellar spin. Our results are
validated by the general torque-dissipation relation in the presence of the stellar
obliquity. We also run the fiducial model in \citet{Laine} and find that the
planet's radius is sustained at a nearly constant value by the Ohmic heating,
rather than being thermally expanded to the Roche radius as suggested by the
authors.
\end{abstract}

\section{Introduction}
In the study of the orbital distribution of known Jupiter-mass exoplanets, the
radial-velocity method has revealed a pile-up of hot Jupiters with orbital periods
of $\sim$ 3
days (e.g., see The Extrasolar Planets Encyclopedia website at
http://exoplanet.eu). A number of models have been proposed to explain the pile-up,
such as an inner disk cavity stopping planet migration
\citep[e.g.][]{Lin96,Rice,Benitez}, and tidal circularization of a gas giant planet
in an extremely eccentric orbit arising from planet-planet interactions after the
proto-planetary disk disperses \citep[e.g.][]{Chatterjee,Nagasawa,WL11,Naoz}. In
addition, tidal heating in a young hot Jupiter in a moderately eccentric orbit may
inflate the planet over its Roche-lobe, resulting in mass loss and therefore
leading to the halting of planet migration or even planet destruction
\citep[e.g.][]{Gu03,Gu04,Chang}. However the excitation of the planet's
eccentricity in this case is subject to the uncertain density profile of the inner
edge of  a disk \citep{Rice,Benitez}.

Aside from the tidal dissipation that relies on the presence of orbital
eccentricity, \citet{Laine} invoked the Ohmic dissipation to inflate a young hot
Jupiter in a circular orbit by adopting the model proposed by Campbell (1983,1997)
for the magnetic interactions in the AM Herculis systems. In this simplified model,
the planet is assumed to behave as an imperfect conductor without its own
ionosphere and magnetosphere; namely, a diamagnetic sphere with a finite
resistivity. In addition, it is assumed that the stellar spin is aligned with the
planet's orbit. Since the vacuum space is assumed between the star and the planet,
the stellar magnetic dipole must be misaligned to induce electric currents and
magnetic fields as the planet circles its T Tauri star. The magnetic torque arises
from the Ohmic dissipation in the planet at the expense of the spin-orbit energy
(see the \S2.3).

Normally a planet even without its own fields possesses an ionosphere due to the
photo-ionization of the upper atmosphere. An induced magnetosphere can form above
the ionosphere \citep{Zhang}. As a planet orbits a star with a tilted magnetic
dipole, the ionosphere may shield the time-varying stellar fields so sufficiently
that little electromagnetic field can be induced in the planet's interior by the
external stellar fields. Nevertheless, one may argue that a young hot Jupiter is
already tidally locked by its parent star such that most of its permanent
night-side lacks an ionosphere. This argument neglects the global circulation in
the atmosphere \citep[e.g.][]{Showman08,Showman09,TC10,RM10,Dobb,Perna}, which may
maintain an ionosphere on the permanent night side. Based on the radio-sounding
results from the Venus Express spacecraft, the ionosphere on the night side of
Venus is weaker and possibly more sporadic than that on the day side
\citep{Patzold}. Hereafter, we boldly apply the model for AM Her binaries to the
entire planet and also consider a vacuum space outside the planet and the star to
simplify the calculation. The consequence of this simplification is that other
electromagnetic effects such as unipolar induction \citep{GL69,LL12}, Alfven-wave
wings \citep{Neubauer,Kopp}, dynamical friction \citep{Papa}, stellar winds
\citep{Vidotto}, planetary winds \citep{Adams,Trammell}, stellar fields diffusing
into the planet interior \citep{Campbell05}, and magnetic reconnections are all
ignored (also see Lanza 2011 for a recent review). In addition, any electromagnetic
effects associated with atmospheric circulations are not being considered for
further simplicity \citep{Perna,BS10,BSB,WL12}. In short, we restrict ourselves
only to the diamagnetic part of the star-planet magnetic interaction\footnote{The
same concept has been applied to star-disk magnetic interactions \citep{Lai99} in
which the magnetic response of the disk is modelled by a diamagnetic disk as well
as a magnetically threaded disk. Unlike our model planet possessing a finite
resistivity, the disk is assumed to be a perfect conductor in the diamagnetic part
of their model.}, as was modelled by \citet{Laine}. It should be noted that the
Ohmic heating proposed by \citet{Laine} is short-lived, since the stellar magnetic
fields decay significantly during the T Tauri phase. This is in contrast with the
Ohmic dissipation model by \citet{BSB}, which is long-lived. Consequently, this
study is concerned exclusively with the early evolution of hot-Jupiter systems.

The observations using the Rossiter-McLaughlin effect \citep{Ohta} suggest that
dwarf stars hosting transiting planets may have possessed a wide range of stellar
obliquities \citep[e.g.][]{Winn10,Winn11}. These observational findings seem
against the conventional paradigm in which a planet should orbit in the same
direction as the stellar spin as the star and planets form together in a
proto-planetary disk. A number of N-body numerical simulations demonstrated that
after the proto-planetary disk disperses, planet-planet interactions accompanied by
tidal circularization, as mentioned in the first paragraph of the Introduction, can
generate obliquities. It was also proposed that before the proto-planetary disk
dissipates, the warp torque resulting from the magnetic interactions between the
proto-star and the inner part of the disk would move the stellar spin away from the
disk angular momentum despite the presence of gas accretion onto the proto-star
\citep{Lai11,FL11}. Motivated by the latter works, it is timely to consider a more
complex case in which stellar obliquity $\lambda$ is not zero; i.e., the orbital
axis is not aligned with the stellar spin.

It should be noted that the tidal dissipation in the star drives the system to the
spin-orbit alignment as well as synchronization \citep[e.g.][]{Hut,Soko,Lai11}. To
make the problem tractable, we do not take account of the influence on $\lambda$
driven by the proto-planetary disk or by tidal interactions with the proto-star,
but simply take $\lambda$ as a free parameter in this work. In addition, we assume
that the planet spin is tightly being synchronized with its orbital motion during
the evolution, therefore generating negligible dissipation in the planet
\citep[e.g.][]{Gu03}. This simplification allows us to ignore the effect due to the
planet spin in the calculation.

Owing to the Ohmic dissipation and the resulting magnetic torques, the stellar
spin, planet's orbit, and the interior structure of the planet evolve
simultaneously. To calculate the coupled evolution more precisely, we adopt an
interior-structure model \citep[for details see][]{Chang} to compute the planet
resistivity and the thermal response of the planet due to the Ohmic heating.

The structure of the paper is organized as follows. In \S2, we describe the
equations for the coupled secular evolutions of stellar spin, planet's orbit, and
planet interior structure due to the diamagnetic interaction between a young hot
Jupiter and its parent T Tauri star. In order to understand the dependence of Ohmic
dissipation on various orientations of the stellar spin and magnetic dipole moment,
we first conduct a parameter study in \S3 to investigate this with no secular
evolutions. The parameter study involving secular evolutions is then presented in
\S4. Finally, we summarize and discuss the results in \S5.

\section{Governing equations for the coupled evolutions of spin, orbit, and planet's interior structure}
\begin{figure}
\epsscale{.70} \plotone{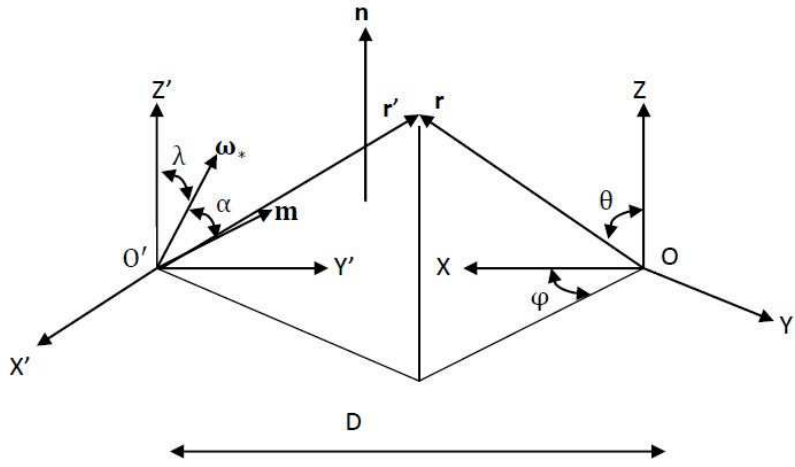} 
\caption{The coordinate systems
adopted in this study for calculations. The $Oxyz$ coordinate system is anchored at
the planet, with the origin $O$ at the planet's center and $x$-axis pointing to the
proto-star. On the other hand, the origin $O'$ of the $O'x'y'z'$ system lies at the
proto-star's center with the $y'$-axis pointing to $O$. The coordinate systems are
the same as those in Laine et al. (2008). At $t=0$ the stellar spin
$\mbf{\omega}_*$ is placed on the $y'$-$z'$ plane and is inclined at the obliquity
angle $\lambda$ relative to the direction of the orbital axis $\bf n$ (i.e. the
vertical axis $z'$). As is viewed in the $Oxyz$ system co-moving with the planet,
$\mbf{\omega}_*$ rotates around the vertical axis with the angle $\lambda$ at the
rate of the orbital angular frequency $n$. Moreover, as the star spins, the stellar
dipole moment $\bf m$ rotates around $\mbf{\omega}_*$ with the misaligned angle
$\alpha$. We let the $y'$ axis always secularly follow the precession of
$\mbf{\omega}_* $ such that $\mbf{\omega}_*$ always lies on the $y'$-$z'$ during
the secular evolution.
\label{fig1}}
\end{figure}

Following the same mathematical procedures in \citet{Laine}, we solve the resistive
induction equation in the co-moving frame of the planet, with the stellar dipole
fields and the induced fields expressed in terms of the poloidal scalars
$\phi_*({\bf r},t)$ and $\phi_p({\bf r},t)$, respectively: namely, the magnetic
field {\bf B}, which has a poloidal nature in our problem, is related to the
poloidal scalars by {\bf B} $= \nabla \times ( \nabla \times (\phi \hat {\bf r}))$,
where $\hat {\bf r}$ is the unit vector of {\bf r}. The SI unit system is adopted
to present the equations for electromagnetic calculations. The induced poloidal
scalar $\phi_p$ can be solved by the separation of variables in the spherical
coordinates ($r$, $\theta$, $\varphi$) of such a frame after $\phi_*$ and the
resistivity profile $\eta(r)$ are given. Let $R_p$ be the planet radius. For
notation convenience, we denote $\phi_p(r>R_p)\equiv \phi_{p,out}$ and
$\phi_p(r<R_p)\equiv \phi_{in}$. Hence the total poloidal scalar outside of the
planet is $\phi_{out}=\phi_*+\phi_{p,out}$. In the case of $\lambda=0$, $\phi_*$
and therefore $\phi_p$ vary at the rate equal to $\omega_*-n$ as viewed by the
planet, where $\omega_*$ is the stellar spin angular frequency and $n$ is the
orbital angular frequency of the planet. In the Appendix A, we illustrate the
coordinate systems for the problem (see Figure \ref{fig1}) and derive the detailed
equations to solve for the potential scalar induced by a tilted magnetic dipole in
the presence of stellar obliquity. We show that in order to describe the
time-varying potential, there will be 3 more frequencies involved other than
$\omega_-\equiv \omega_* -n$; they are $\omega_+\equiv \omega_* +n$, $\omega_*$ and
$n$. Once the potential scalar $\phi_p$ is solved, the induced magnetic field
$=\nabla \times ( \nabla \times \phi_p {\bf \hat r} )$, the electric field $\bf E$,
the electric current $\bf j$, and hence the Ohmic dissipation can be all calculated
\citep{Laine}. In the following subsections, we describe how to calculate the
resistivity of the planet and the corresponding spin and orbital evolutions due to
the Ohmic dissipation in our model.

\subsection{Calculation of resistivity}


For a hot Jupiter around a T Tauri star of one solar luminosity, the equilibrium
temperature is $\sim 1000-2000$ K at the photosphere. The gas in the region just
below the photosphere is therefore weakly ionized due to thermal ionization of
alkaline elements. As the temperature and density continue to rise in deep layers,
the thermal ionization of the most abundant constituents H and He starts to become
non-negligible. In the even deeper interior, the density is high enough so that the
fluid is partially degenerate and fully ionized due to pressure ionization
\citep{Saumon}. It has been shown that the electric currents and magnetic fields
induced near the planet's surface are only present in the outer part of the planet
where the ionization fraction is low and hence the resistivity is high. That is,
magnetic fields decrease significantly over a skin depth $\delta$ from the surface
to the interior \citep{Laine,BS10}. In other words, the induced electric currents
and magnetic fields are considerably shielded out by the outer part of the planet
such that the precise values of resistivity in the interior do not matter. Thus in
this work, we restrict ourselves to the resistivity $\eta$ due to electron-neutral
collisions in a weakly ionized plasma \citep{Draine,BB94}:
\begin{equation}
\eta_{e-n}=230 \left( {n_n\over n_e } \right) T^{1/2} \ {\rm cm^2/s},\label{eq:eta}
\end{equation}
and apply the above equation to the entire planet without making a significant
error. In the above equation, $n_n$ is the neutral number density, $n_e$ is the
electron number density, and $T$ is the temperature.


To estimate the ionization fraction in eq.(\ref{eq:eta}), we first consider the
thermal ionization of alkaline elements. Thermal ionization is governed by the Saha
equation \citep[cf.][]{BB94,Perna}
\begin{equation}
{n_e \over n_n} \approx  {1\over n_n^{1/2}} \left( {m_e k T\over 2\pi \hbar^2}
\right)^{3/4} \sqrt{\sum_j f_j \exp{(-I_j/kT)}}, \label{eq:ionize1}
\end{equation}
where $n_e=\sum n_j^+$, $n_j=f_j n$, $k$ is the Boltzmann's constant, $\hbar$ is
the Planck's constant divided by $2\pi$, and $n_j^+ \ll n_j$ is assumed. We follow
\citet{BS10} to find the abundances $f_j$ and ionization potential $I_j$ of each
alkaline species (labelled by $j$) inferred from \citet{Lodders}\footnote{$f_j$ are
estimated at the temperature $> 2000$ K. Below this temperature, the abundances of
some species such as Fe and Ca decline dramatically due to their molecular
formations with other atoms. This process does not affect our results significantly
because K and Na are the primary sources of thermal electrons at the low
temperatures.}
and \citet{Cox}.

In even deeper layers, the thermal ionization of the most abundant constituents H
and He starts to dominate the electron contribution. We compute the H \& He
ionization based on the equation of state tables in \citet{Saumon}. Given the
pressure $P$, temperature $T$, and the helium mass fraction $Y$, the mass density
is given by \citep{Saumon}
\begin{equation}
{1\over \rho(P,T)}={1-Y \over \rho^H(P,T)}+{Y\over \rho^{He}(P,T)},
\end{equation}
where $\rho^H(P,T)$ and $\rho^{He}(P,T)$ are obtained from interpolation of the
data in the EoS tables for pure H and He, respectively. Hence, the electron number
density $n_e$ and the total number density $n$ are given by (see eqs.(36) \& (37)
in Saumon et al. 1995)
\begin{eqnarray}
n_e&=&n_e^{H}+n_e^{He}={2\rho_H/m_H \over 1+3X_{H_2}+X_H}X_e^H +{3\rho_{He}/m_{He}
\over 1+2X_{He}+X_{He^+}}X_e^{He}, \label{eq:ionize2}\\
n&=&n_{H_2,H,H^+,e}+n_{He,He^+,He^{2+},e}={2\rho_H/m_H \over
1+3X_{H_2}+X_H}+{3\rho_{He}/m_{He} \over 1+2X_{He}+X_{He^+}},\label{eq:ionize3}
\end{eqnarray}
where $\rho_H\approx \rho (1-Y)$, $\rho_{He}=\rho-\rho_H=Y\rho$, $X_e^H$ and
$X_e^{He}$ are given by eqs.(34) and (35) respectively in \citet{Saumon}. $Y=0.283$
is adopted in our interior-structure simulations.


\subsection{Planet radius and spin-orbital evolution due to Ohmic heating}

Including the Ohmic dissipation but neglecting the small planetary spin energy
\citep{Bodenheimer}, we have
 the evolution of the global energy for the entire planet governed by \citep[cf.][]{Chang}
\begin{equation}
\dot U + \dot W  = \dot Q_{ohmic} - L,\label{eq:energy}
\end{equation}
where $U$ is the internal energy, $W$ is the gravitational potential energy, $L$ is
the intrinsic luminosity from the photosphere of the planet, and $\dot Q_{ohmic}$
is the Ohmic dissipation rate given by \citep{Laine}
\begin{equation}
\dot Q_{ohmic}=
\left\langle \int_{r\leq R_p} \mu_0 \eta {\rm Re}({\bf j})^2 dV \right\rangle
=\left\langle \int_{r\leq R_p} {\eta \over \mu_0} [{\rm Re}(\nabla \times {\bf
B})]^2 dV \right\rangle, \label{eq:ohmic_heat}
\end{equation}
where ``Re" means taking the real part and $\langle \rangle$ denotes the time
averaging over the time scale longer than the forcing periods; namely, it is the
secular evolutions of the spin and orbit that are relevant to the long-term thermal
evolution of the interior structure. $\dot Q_{ohmic}$ should be equal to the
average flow of the electromagnetic power (i.e. the Poynting vector) into the
planet through the planet's photosphere \citep[e.g.][]{Jackson}.

Owing to the diffusive nature of the problem, an order-of-magnitude estimate for
the Ohmic dissipation can be made based on the dimensional analysis of Equation
\ref{eq:ohmic_heat} with $\nabla \sim 1/\delta$ and $dV \sim R_p^2 \delta$
\citep{Campbell97,Laine}
\begin{equation}
 \dot Q_{ohmic} \sim {B^2
\over 2\mu_0} (4\pi R_p^2 \delta) \omega,\label{eq:skin_depth}
\end{equation}
where the skin depth for our magnetic induction problem is $\delta =
(2\eta/\omega)^{1/2}$, the stellar magnetic field near the planet is $B \sim
(\mu_0/4\pi)m/D^3$, $\omega$ is the forcing frequency, $m$ is the magnitude of the
stellar dipole moment, and $D$ the orbital separation (see Figure~\ref{fig1}).

The spin-orbit evolution is dictated by the dissipation torque. The torque acting
on the stellar spin due to the electromagnetic interaction in the inertial frame
takes the form \citep[cf.][]{Campbell97}
\begin{equation}
{\bf T_{inert}=m_{obliquity,inert} \times B_{planet}(r=r_{star})}={\bf
m_{obliquity,inert}}\times \nabla (\partial_r \phi_p)_{\bf r=r_{star}}.
 \end{equation}
In the above equation, $\bf m_{obliquity,inert}$, given by ${\bf P_{x'} m}$ (see
the Appendix A), is the stellar dipole moment as seen in the inertial frame with
{\bf n} in the $z$-direction, and ${\bf B_{planet}}$ is the planet-induced magnetic
field at the location of the star ${\bf r_{star}}=D{\bf \hat i}=D{\bf \hat
r}+(\pi/2)\mbf{\hat \theta}$ (see Figure~1). Note that although ${\bf B_{planet}}$
is calculated in the planet's rest frame, its value in the inertial frame is the
same as in the non-relativistic regime when the terms of order $n^2D^2/c^2 \ll 1$
are neglected, where $c$ is the speed of light \citep[e.g.][] {TM09}.

We shall see that the stellar spin precesses secularly with time in the inertial
frame and we are interested in the spin-orbit evolution on the secular timescale,
much larger than the spin and orbital periods. This amounts to taking the
time-average for each physical quantity to average out their short-term variations.
Besides, it is convenient to work out the secular evolution problem in the
precession coordinates with {\bf n} always pointing to the $z$-direction. Hence
after each time step, we switch to
the inertial frame such that the stellar spin is always on
the $y'$-$z'$ plane and the orbital angular momentum is always along the $z'$-axis
at the beginning of the next time step. We denote this ``instantaneous" inertial
frame as $O'x''y''z'$, which coincides with $O'x'y'z'$ at the beginning of each
time step.
Thus the time-averaged torque $\langle \bf T_{inert} \rangle$ in this inertial
frame is given by \citep[cf.][]{Campbell97}
\begin{eqnarray}
\langle T_{x''} \rangle &=& m\langle \hat m_{z'} B_{planet,r}-\hat m_{y''} B_{planet,\theta} \rangle,\\
\langle T_{y''} \rangle &=& m\langle \hat m_{x''} B_{planet,\theta}+\hat m_{z'} B_{planet,\varphi} \rangle,\label{eq:T_y}\\
\langle T_{z'}\rangle &=& -m\langle \hat m_{x''} B_{planet,r}+\hat m_{y''}
B_{planet,\varphi} \rangle,\label{eq:T_z}
\end{eqnarray}
where $\hat m_{x''}$, $\hat m_{y''}$, and $\hat m_{z'}$ are the three Cartesian
components of the unit vector of $\bf m_{obliquity,inert}$.

Once the time-averaged torque is obtained in the inertial frame, we are ready to
calculate the secular evolutions of spin and orbit. First of all, $\langle T_{x''}
\rangle$ leads to the precession of stellar spin around the orbital axis; namely,
\begin{equation}
f(\lambda) I_* \omega_* {d \langle \varphi'' \rangle \over dt} = \langle T_{x''}
\rangle,\label{eq:precession}
\end{equation}
where $\langle \varphi'' \rangle$ refers to the time-averaged precession angle and
$f(\lambda)$ is a function of $\lambda$ given in the Appendix B. In the
``instantaneous" inertial frame, $\langle {\bf T_{inert}} \rangle$ along the
stellar axis determines $\dot \omega_*$. $-\langle T_{z'} \rangle$ governs $\dot n$
and thus $\dot D$. Moreover, $\dot \lambda$ is caused by the components of $\langle
T_{y''}\rangle$ and $\langle T_{z'} \rangle$ normal to the stellar spin, and
additionally by the back reaction $-\langle T_{y''}\rangle$ acting to the orbital
angular momentum. In other words, using the ``instantaneous" stellar spin
$\mbf{\omega}_*=\omega_* \mbf{\hat \omega}_*=\omega_*(\sin \lambda {\bf \hat j''} +
\cos \lambda {\bf \hat k'})$, we arrive at a set of evolutionary equations:
\begin{eqnarray}
&&{d(I_* \omega_*)\over dt}= \langle {\bf T_{inert}} \rangle \cdot \mbf{\hat\omega}_*=\langle T_{y''}\rangle \sin \lambda + \langle T_{z'}\rangle \cos \lambda, \label{eq:ode1}\\
&&{d\lambda \over dt}={\langle T_{y''}\rangle \cos \lambda - \langle T_{z'}\rangle
\sin \lambda \over I_*
\omega_*}+{\langle T_{y''}\rangle \over M_p D^2 n},\label{eq:T_lambda}\\
&&M_p {d(D^2 n) \over dt} = -\langle T_{z'}\rangle \label{eq:ode4},
\end{eqnarray}
where $M_p$ is the planet's mass and the complicated expression for $d\lambda /dt$
is explained in the Appendix B. The above 3 equations can be combined to express
$\dot \lambda$ in terms of $\dot L_{spin}=d(I_*\omega_*)/dt$ and $\dot
L_{orb}=M_pd(D^2 n)/dt$ as follows
\begin{equation}
{d\lambda \over dt}={dL_{orb}\over dt} \left( {1\over  I_* \omega_* \sin \lambda} +
{1\over  M_p D^2 n\tan \lambda } \right) + {d L_{spin} \over dt} \left( {1\over I_*
\omega_*\tan \lambda } + {1\over M_p D^2 n\sin \lambda }
\right),\label{eq:T_lambda1}
\end{equation}
which we shall find quite useful to interpret the evolutionary results.

Note that the moment of the inertia $I_*$ of the T Tauri star also evolves. In
reality, $m$ evolves as well \citep{JK,YJ11}, but in this work we prescribe a
constant value for $m$ for simplicity.

When $\lambda=0$, the terms in the magnetic potential scalar $\phi_p$ associated
only with $\omega_-$ are left, leading to $\langle T_{x''} \rangle=\langle
T_{y''}\rangle=0$. Therefore Equations (\ref{eq:ode1})-(\ref{eq:ode4}) reduce to
the ones in \citet{Campbell83}:
\begin{eqnarray}
{d(I_* \omega_*)\over dt}&=&  \langle T_{z'} \rangle, \\
 M_p {d(D^2 n) \over dt} &=& -\langle T_{z'} \rangle.
\end{eqnarray}

In the absence of stellar obliquity, the torque and Ohmic heating can be simply
related to each other by virtue of the equation $\dot Q_{ohmic}=|\omega_- \langle
T_{z'} \rangle |$ \citep{Campbell83,Laine}. If $\lambda \neq 0$, we show in the
\S2.3 that the vector product ${\bf \langle T_{inert} \rangle} \cdot
(\mbf{\omega}_* -{\bf n})$ gives rise to the Ohmic dissipation that drives the
spin-orbit system toward a lower energy state.
More specifically,
\begin{eqnarray}
\dot Q_{ohmic} &=& -\left( \omega_* \langle T_{y''} \rangle \sin \lambda + \omega_*
\langle T_{z'} \rangle \cos \lambda  -n\langle T_{z'} \rangle
\right),\label{eq:heat_torque}\\
&=& -\omega_* {dL_{spin}\over dt}+n{dL_{orb}\over dt}.
\end{eqnarray}
Since $\dot Q_{ohmic}$, $\omega_*$, and $n$ are all positive quantities in this
work, the above equation indicates that $dL_{spin}/dt > dL_{orb}/dt$ for $\omega_*
<n$ and vice versa, which is a familiar result for $\lambda=0$ but even applies
generally to the cases for $\lambda \neq 0$. Note that even when $n=\omega_*$,
$\dot Q_{ohmic} \neq 0$ due to the spin-orbit misalignment. In addition, $\langle
T_{x''} \rangle$ causing the precession of stellar spin around the orbital axis
does not do any mechanical work and thus is not related to the Ohmic dissipation in
the planet.

In the special case where $\alpha=0$, the stellar dipole moment in the inertial
frame is ${\bf \hat m_{obliquity,inert}}=(0,\sin\lambda,\cos\lambda)$.
Substituting this into eqs.(\ref{eq:T_y}) \& (\ref{eq:T_z}), we have the unique
relation $\langle T_{z'}\rangle/ \langle T_{y''} \rangle =-\tan \lambda$ regardless
of the value of $\omega_*$. This together with eq.(\ref{eq:heat_torque}) gives
\begin{equation}
\dot Q_{ohmic}=n\langle T_{z'} \rangle,\label{eq:alpha=0}
\end{equation}
which is independent of  $\omega_*$ as it should be when the spin and stellar
dipole are aligned. Once again, $\dot Q_{ohmic} >0$ and $n>0$ by our sign
convention. It follows from the above equation that $\langle T_{z'}\rangle >0$,
therefore always leading to an orbit decay.

\subsection{General relation between energy dissipation and torques} Since the
torques arise from energy dissipation, we wish to derive the relation between the
torques and dissipation in the presence of obliquity. This relation provides a
powerful check on whether the Ohmic dissipation and the resulting torques
calculated in \S2.2 are correct.

The stellar spin angular momentum and planet's orbital orbital angular moment are
given by
\begin{equation}
{\bf L_{spin}}= I_* \mbf{\omega}_*,
\end{equation}
and
\begin{equation}
{\bf L_{orb}}=(m_p \sqrt{Gm_* a} ){\bf \hat n},\label{eq:L_orb}
\end{equation}
respectively. Since the total angular momentum is conserved, ${\bf T''}=d{\bf
L_{spin}}/dt=-d{\bf L_{orb}}/dt$.

However the total energy of the system is not conserved as a result of dissipation.
The stellar spin energy changes at a rate according to
\begin{equation}
{dE_{spin}\over dt}={d (1/2) I_* \mbf{\omega}_*^2 \over dt} = \mbf{\omega}_* \cdot
{ d I_* \mbf{\omega}_* \over dt} = \mbf{\omega}_* \cdot {d {\bf L_{spin}}\over dt}.
\end{equation}
The change rate of the orbital energy is
\begin{equation}
{dE_{orb} \over dt}={GM_*m_p\over 2 a^2}{da\over dt},
\end{equation}
which can be linked to the change of the orbital angular momentum as follows
\begin{equation}
{dE_{orb} \over dt}={d{\bf L_{orb}}\over dt} \cdot {\bf n}.
\end{equation}
In deriving the above equation, we have taken the time derivative of
Equation~(\ref{eq:L_orb}) and used the identity $d {\bf \hat n}/dt \cdot {\bf
n}=0$. Thus
\begin{equation}
-\dot Q_{ohmic}\equiv {d(E_{spin}+E_{orb})\over dt}=\mbf{\omega}_* \cdot {d {\bf
L_{spin}}\over dt}+{d{\bf L_{orb}}\over dt} \cdot {\bf n}={\bf T''} \cdot
(\mbf{\omega}_* -{\bf n}).
\end{equation}
Although the heating rate is expressed in terms of the Ohmic dissipation, the above
relation for dissipative torques can apply generally to other dissipative processes
such as tidal dissipation. It is apparent that we do not specify the Ohmic
dissipation to deduce the above relation.

\subsection{Summary of procedures} Given $I_*(t)$ and $M_p$, the evolutions of stellar spin and
orbit ($\omega_*$, $\lambda$, and $D$) are coupled with the evolution of interior
structure ($R_p$, $\eta(r)$ etc.) via the Ohmic dissipation in a hot Jupiter, which
is modelled as a diamagnetic sphere in our calculation.

The procedure of the evolutionary calculations is summarized as follows. We start
with initial $I_*$, $\omega_*$, $\lambda$, $D$, and $R_p$ to obtain $\dot
Q_{Ohmic}$ and ${\bf T_{inert}}$.
The next step consists of three calculations: the first is the calculation of the
new interior structure of the young hot Jupiter due to $\dot Q_{Ohmic}$,
the second is the computation of the new $I_*$ from a stellar code,
and the last is the calculation of the new $\omega_*$, $\lambda$, $D$ from the
integration of the ODEs from Equation~(\ref{eq:ode1}) to Equation~(\ref{eq:ode4})
based on ${\bf T_{inert}}$. Consequently, $I_*$, $\omega_*$, $\lambda$, $D$, and
$R_p$ at the next time step will be obtained. Meanwhile, the computed $\bf
T_{inert}$ and $\dot Q_{Ohmic}$ can be checked using
Equation~(\ref{eq:heat_torque}) to validate the calculation. The same procedure is
then carried out over and over again to evolve the system until either the planet
reaches its Roche radius or the calculation approaches the end of simulation at
$10^7$ years. We employ the same codes described and used by \citet{Bodenheimer}
and \citet{Chang}
for the planetary and stellar interior structures, respectively.

 To simulate the stellar rotation being locked by a process such as disk locking
\citep[e.g., see][and reference therein]{Chang}, we also run cases (actually most
of the cases) in which the stellar spin $\omega_*$ is held at its initial value
throughout the simulation, even though $\lambda$ is still allowed to evolve by the
magnetic interaction. It is conceivable that any external torques affecting
$\omega_*$ should change $\lambda$ as well, such as the star-disk magnetic
interaction by \citet{Lai11} and \citet{FL11}. In this study, the star-disk
interaction is not modelled with the star-planet magnetic interaction. Instead, we
focus only on the evolutions due to the star-planet magnetic interaction, with the
condition for $\omega_*$ to be ``locked" for the sake of simplicity of the toy
model.

\section{Comparative studies without secular evolutions}
In this section, we present a couple of test runs in our model without considering
the evolution of spin, orbit, and interior structures of the proto-star and planet.
The purpose of the test runs is to investigate how the Ohmic dissipation varies
with $\lambda$ and $\alpha$. This provides parameter and thus comparative studies
to understand the basic behavior of the results before we proceed to the more
complicated calculations involving secular evolutions.

\begin{figure}
\includegraphics[scale=.45,angle=-90]{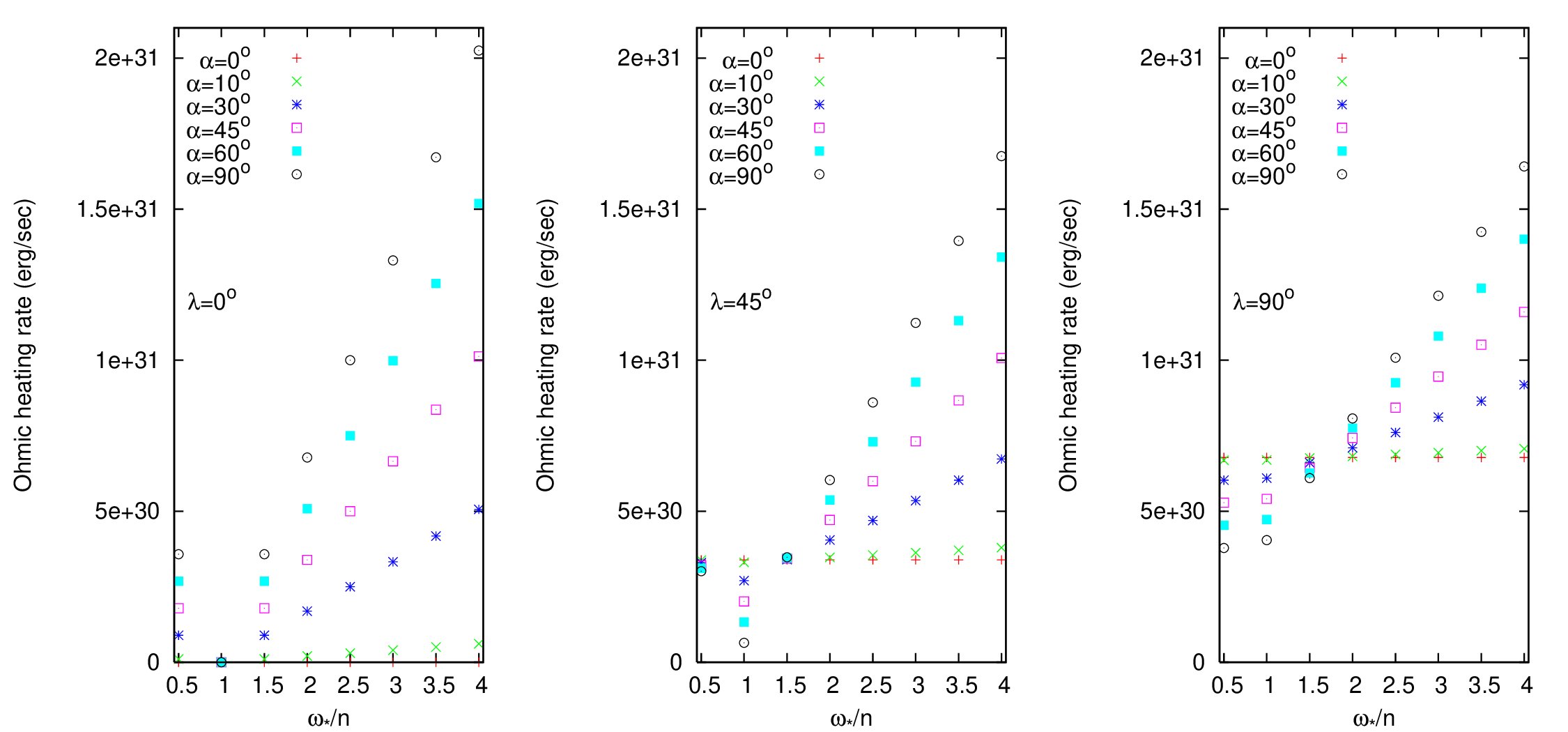}
\includegraphics[scale=.45,angle=-90]{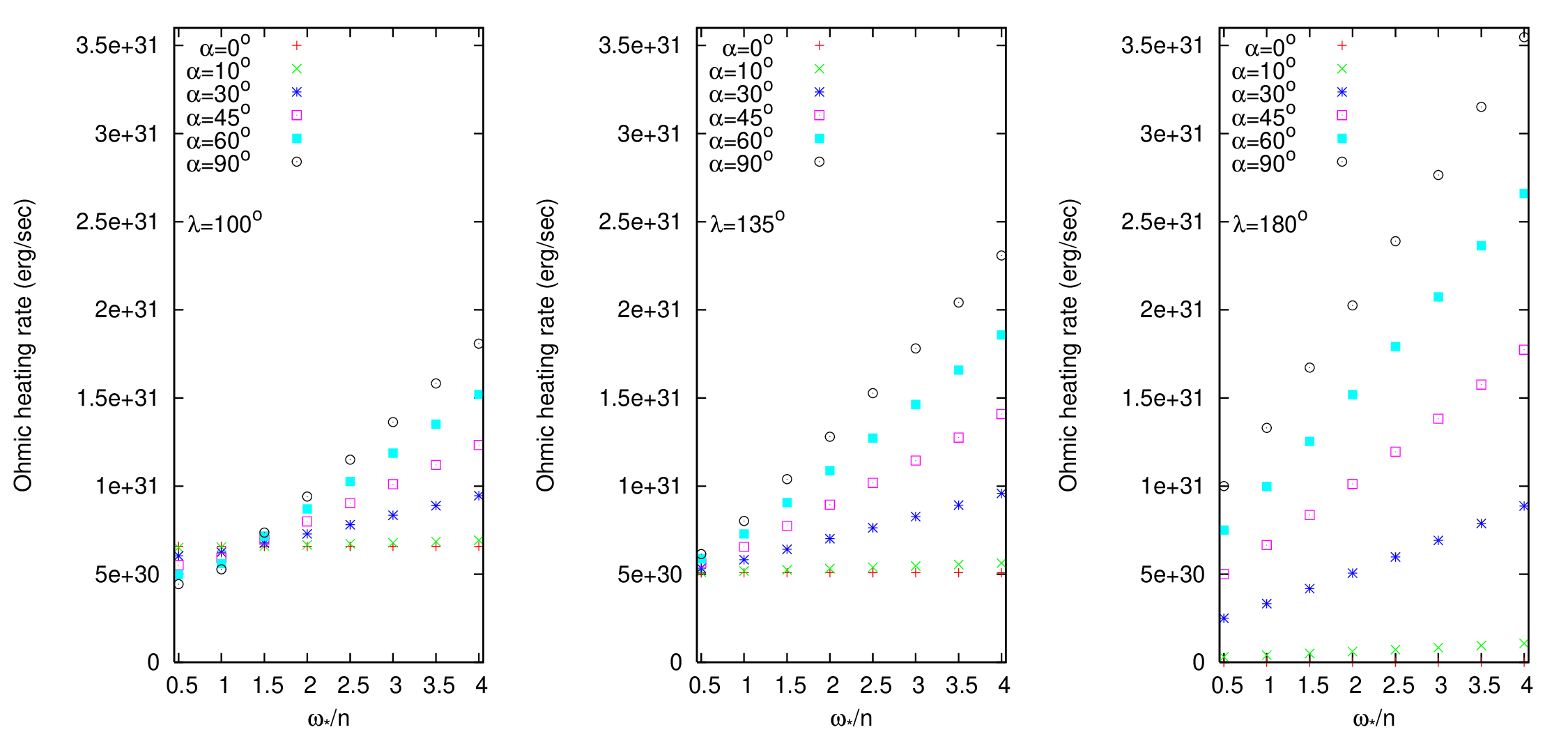}
 \caption{Dependence of the Ohmic heating rate on $\lambda$, $\alpha$, and
$\omega_*/n$. The Ohmic heating rates are calculated from
Equation(\ref{eq:ohmic_heat}). The mass, radius, and orbital radius of the young
hot Jupiter are 1 Jupiter mass, 1.84 Jupiter radii, and 0.02 AU for this set of
calculations.\label{fig2}}
\end{figure}

Such comparative studies for the Ohmic dissipation rate $\dot Q_{ohmic}$ vs.
$\omega_*/n$ are shown in Figure~\ref{fig2}. $\dot Q_{ohmic}$ is calculated by
virtue of Equation~(\ref{eq:ohmic_heat}). We adopt $m=4\times 10^{34}$ A m$^2$, the
same fiducial value used in \citet{Laine}. The interior structure for a coreless
hot Jupiter with $M_p=M_J$ and $R_p=1.84R_J$ is used for the test runs. We consider
$\omega_*$ as a free parameter, whereas $n$ is held constant corresponding to the
orbital radius of 0.02 AU. Thus, the stellar irradiation, which affects the
interior structure, is also constant. This reduces the number of variables and
helps to more easily examine how $\dot Q_{ohmic}$ varies with $\omega_*/n$ in the
test runs. Everything else being the same, there is no difference in the Ohmic
dissipation for $\alpha$ and for ($180^{\circ}-\alpha$) due to the axi-symmetry of
dipole fields. Consequently, we only present the cases for $\alpha \leq 90^{\circ}$
in Figure~\ref{fig2}.

The upper left panel of Figure \ref{fig2} shows that in the absence of the stellar
obliquity ($\lambda=0$), the Ohmic heating rates increase from zero for $\alpha=0$
to the maximum values for $\alpha=90^\circ$. In addition, the heating rate vanishes
when $\omega_{-}=0$ (i.e. $\omega_*/n=1$) and increases with $|\omega_{-}|$ due to
stronger electromagnetic interactions induced by faster forcing. The trend and the
character of these results agree with those in \citet{Laine}. The similar line of
argument applies to the cases for $\lambda=180^{\circ}$ in which the stellar spin
is completely flipped over and therefore the forcing frequency is $\omega_+$ rather
than $\omega_-$. As illustrated in the lower right panel of Figure~\ref{fig2}, the
Ohmic heating rate increases with $\alpha$. Besides, the heating rate increases
with $\omega_*$ and thus $\omega_+$.

We find that the dissipation torque $|T_{z'}|$ for $\lambda=0^\circ$ decreases with
the forcing frequency $|\omega_-|$ except when the forcing frequency is very close
to zero; i.e. the torque peaks at $\omega_-\approx 1.3\times 10^{-4}$ s$^{-1}$ for
$\lambda=0^\circ$. The maximum value of the torque arises because the torque is
significantly weak for extremely slow forcing, and becomes small again for fast
forcing due to the dissipation localized within one small skin depth below the
planet's surface \citep{Campbell83,Campbell97}.

\begin{figure}
\includegraphics[scale=.33,angle=-90]{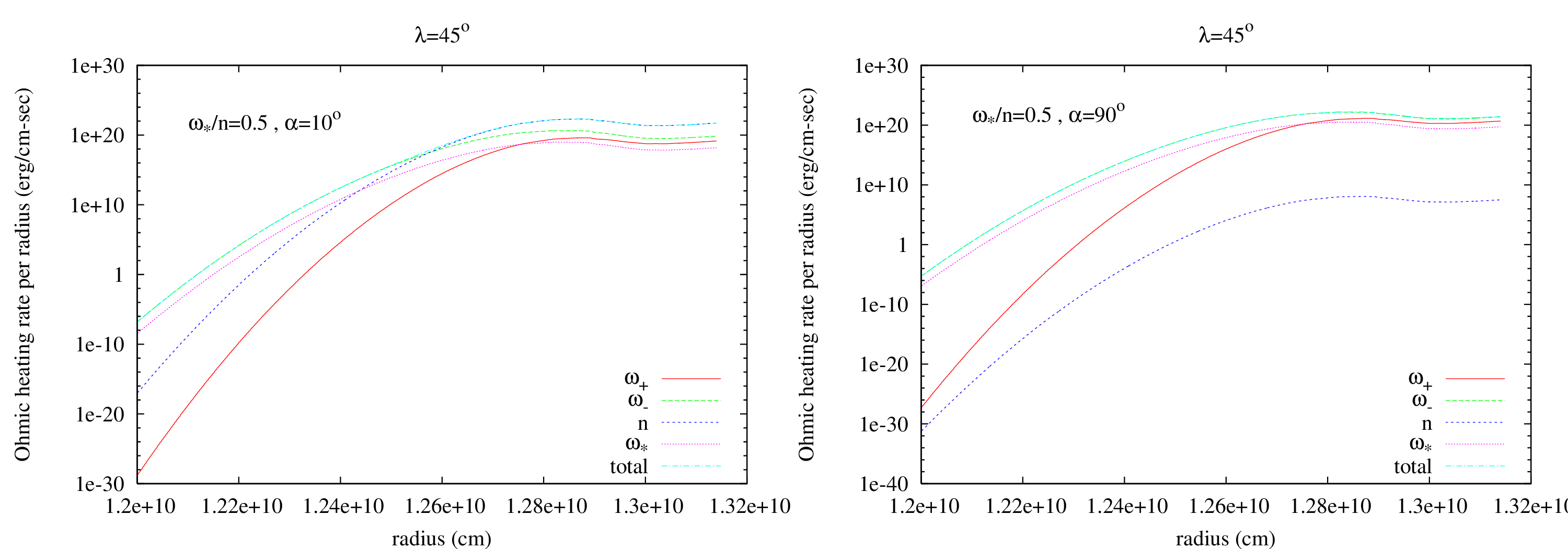}
\includegraphics[scale=.33,angle=-90]{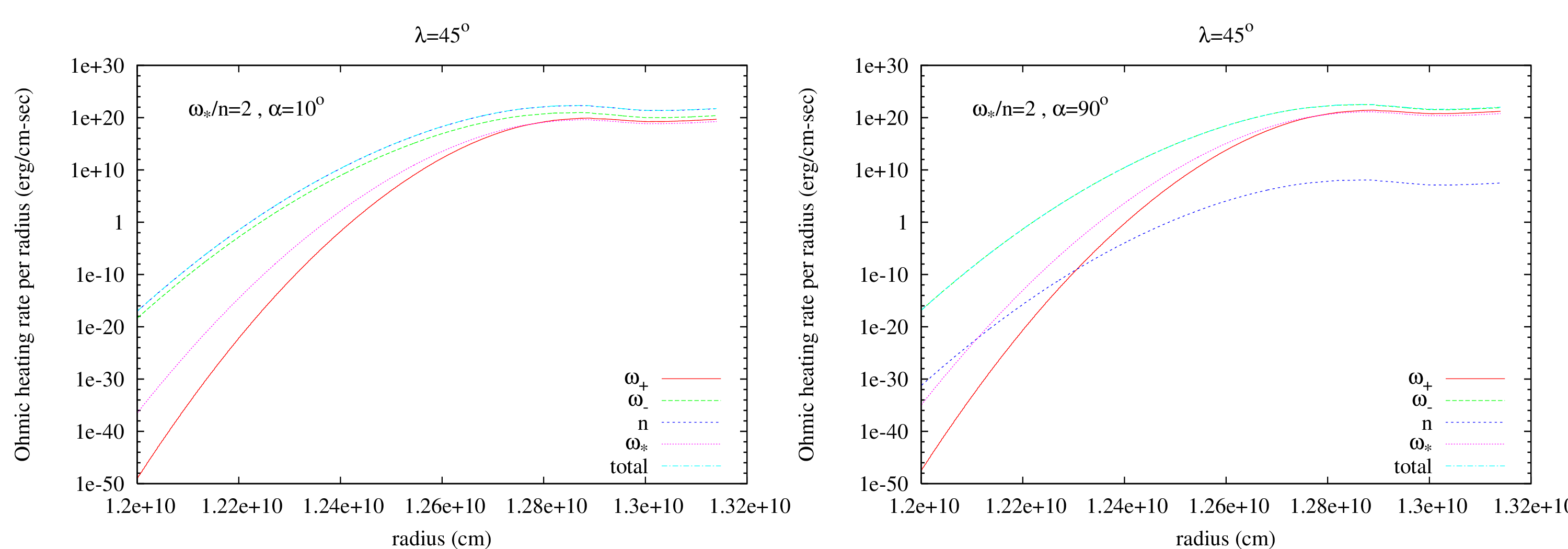}
\caption{Ohmic heating rate profiles in a young hot Jupiter of 1 Jupiter mass and
1.84 Jupiter radii.
$\alpha=10^\circ$ and $90^\circ$ for the stellar obliquity $\lambda =45^{\circ}$
are plotted.
As illustrated in the figure, most of the Ohmic dissipation occurs in the outer
part of the planet because the induced electromagnetic fields can only penetrate
from the surface down over a length scale comparable to the skin depth $\sim
\sqrt{\eta/\omega}$. The radiative-convective interface is located at $\approx
1.295\times 10^{10}$ cm. Note that the total heating profile overlaps with the
profiles associated with other forcing frequencies (see the text in \S3 for the
details). \label{fig5}}
\end{figure}

\begin{figure}
\epsscale{.70}
\includegraphics[scale=.33,angle=-90]{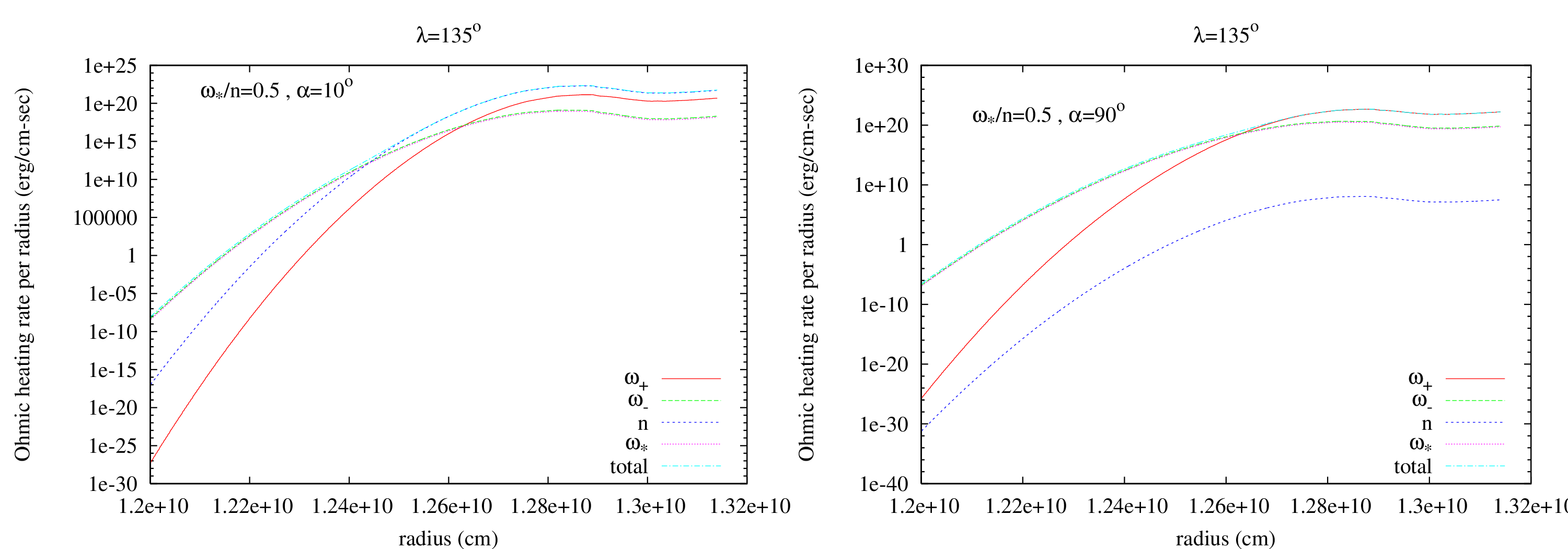}
\includegraphics[scale=.33,angle=-90]{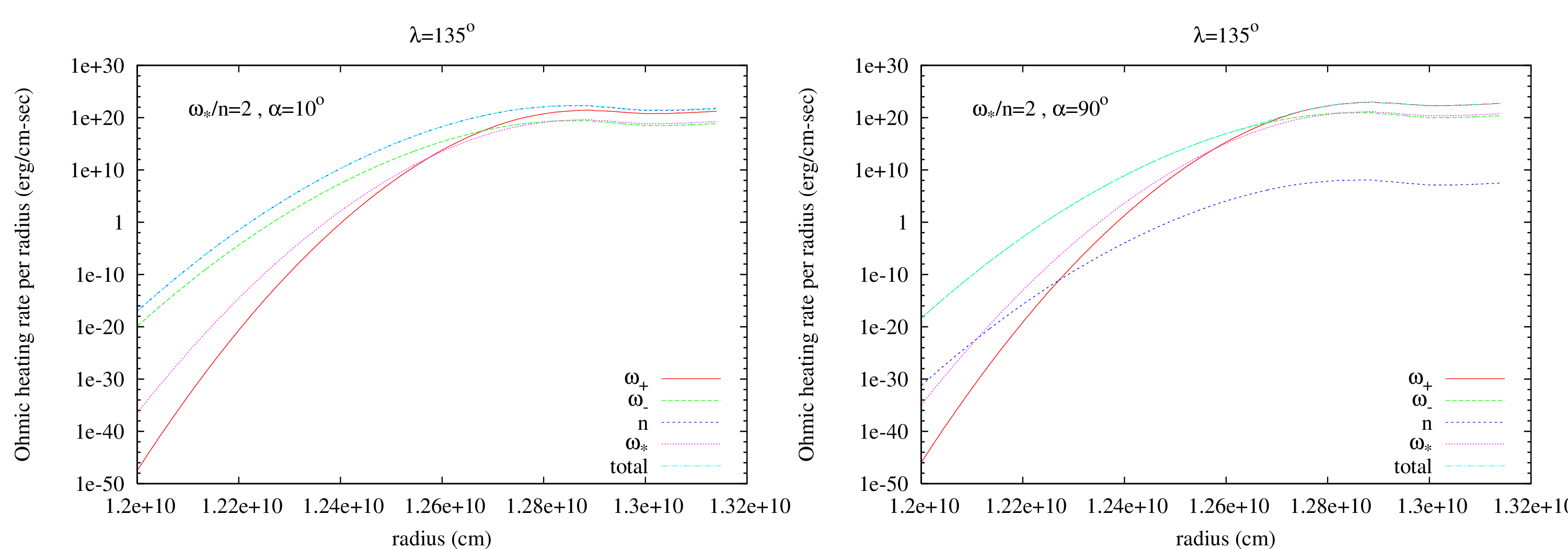}
\caption{Same as Figure \ref{fig5} but for a planet in a retrograde orbit with
$\lambda=135^{\circ}$.\label{fig6}}
\end{figure}

When $\lambda \neq 0$, the relation between $\omega_*$ and $\dot Q_{ohmic}$ becomes
perplexing and requires more explanations. As shown in Figure \ref{fig2} for
$\lambda=45^\circ$, $90^\circ$, and $100^{\circ}$, the positive correlation between
$\omega_*$ and $\dot Q_{ohmic}$ exists when $\omega_*/n$ is large enough for the
forcing frequency $\omega_*$ to play the main role. This outcome can be realized by
contemplating the problem in the two extreme regimes: $\alpha \gtrsim 0$ and
$\alpha \lesssim 90^{\circ}$; the orbital motion alone contributes most of the
heating in the former regime, whereas in the latter regime the Ohmic heating is
generated primarily from the relative spin-orbit motion (i.e. $\omega_-$ or
$\omega_+$ depending on $\lambda$). More specifically, Figure~\ref{fig2} shows that
the heating rate is constant independent of $\omega_*$ for the cases of $\alpha=0$,
in agreement with Equation~(\ref{eq:alpha=0}); namely, the Ohmic dissipation
induced entirely by the orbital motion with the forcing frequency $n$. As $\alpha$
starts to deviate from zero, we find that the Ohmic dissipations induced by other
forcing frequencies begin to increase but the Ohmic heating arising solely from the
orbital motion starts to decrease. This can been seen in Figures~\ref{fig5} \&
\ref{fig6} for $\alpha=10^\circ$ where the total heating profile (shown in cyan
line) near the planet surface almost overlaps with the one corresponding to the
forcing frequency $n$ (dotted blue line), but the heat contributions from other
frequencies other than $n$ are not totally negligible.
When $\alpha=90^{\circ}$, the heat contribution from the forcing frequency
$\omega_+$ or $\omega_-$ as a result of the relative spin-orbit motion totally
dominates over that from the forcing frequency $n$; namely, in the outer part of
the planet, the total heating profile (cyan line) almost coincides with the one for
$\omega_-$ (dashed green line) in Figure~\ref{fig5} and for $\omega_+$ (magenta
line) in Figure~\ref{fig6}. It can be confirmed by Equations~(\ref{eq:m_obliquity})
and (\ref{eq:phi_*}) that when $\alpha=90^\circ$, the forcing with the frequency
$n$ disappears in the expression of the stellar magnetic dipole moment ${\bf \hat
m_{obliquity}}$ and thus in the corresponding poloidal scalar $\phi_*$, leading to
null contribution of the dissipation from the forcing frequency $n$. Therefore, the
tiny dissipations for $\alpha=90^\circ$ shown in Figures~\ref{fig5} \& \ref{fig6}
(i.e. dotted blue line) stem totally from numerical errors, which are too small to
affect the results. Note that the dissipation occurs primarily in the outer part of
the planet because the induced magnetic fields are mostly confined within one skin
depth below the photosphere.

Given the above explanations, we are able to further elaborate the general
dependence of $\dot Q_{ohmic}$ on $\lambda$ and $\alpha$ shown in
Figure~\ref{fig2}. Let's first examine the cases for $\lambda=45^{\circ}$,
$90^{\circ}$ and $100^{\circ}$. In these cases, the heating rate in the high
frequency range $\omega_*/n > 1.5$-2 increases with $\alpha$. Roughly speaking, it
is because the primary forcing switches from the slow rate $n$ to the fast rate
$\omega_+$ (for $\lambda \geq 90^\circ$) or $\omega_-$ (for $\lambda < 90^\circ$)
as $\alpha$ increases (see the $\omega_*/n=2$ case in Figure \ref{fig5}). The trend
apparently reverses in the low frequency range $\omega_*/n < 1.5$-2; namely, the
heating rate decreases with the increasing $\alpha$ (see the $\omega_*/n=0.5$ case
in Figure \ref{fig5}).


In contrast, for the retrograde orbits with large stellar obliquities as
represented by the cases for $\lambda=135^{\circ}$ and $180^{\circ}$ in Figure
\ref{fig2}, the Ohmic dissipation always increases with $\alpha$. The forcing with
the frequency $\omega_+$ is always fast enough to induce more heat for larger
$\alpha$ than the heat generated mostly by the slower forcing with the frequency
$n$ for smaller $\alpha$. This consequence can be implied by comparing the heating
profiles for $\omega_*/n=2$ with those for $\omega_*/n=0.5$ in Figure \ref{fig6}.

\begin{figure}
\includegraphics[scale=.45,angle=-90]{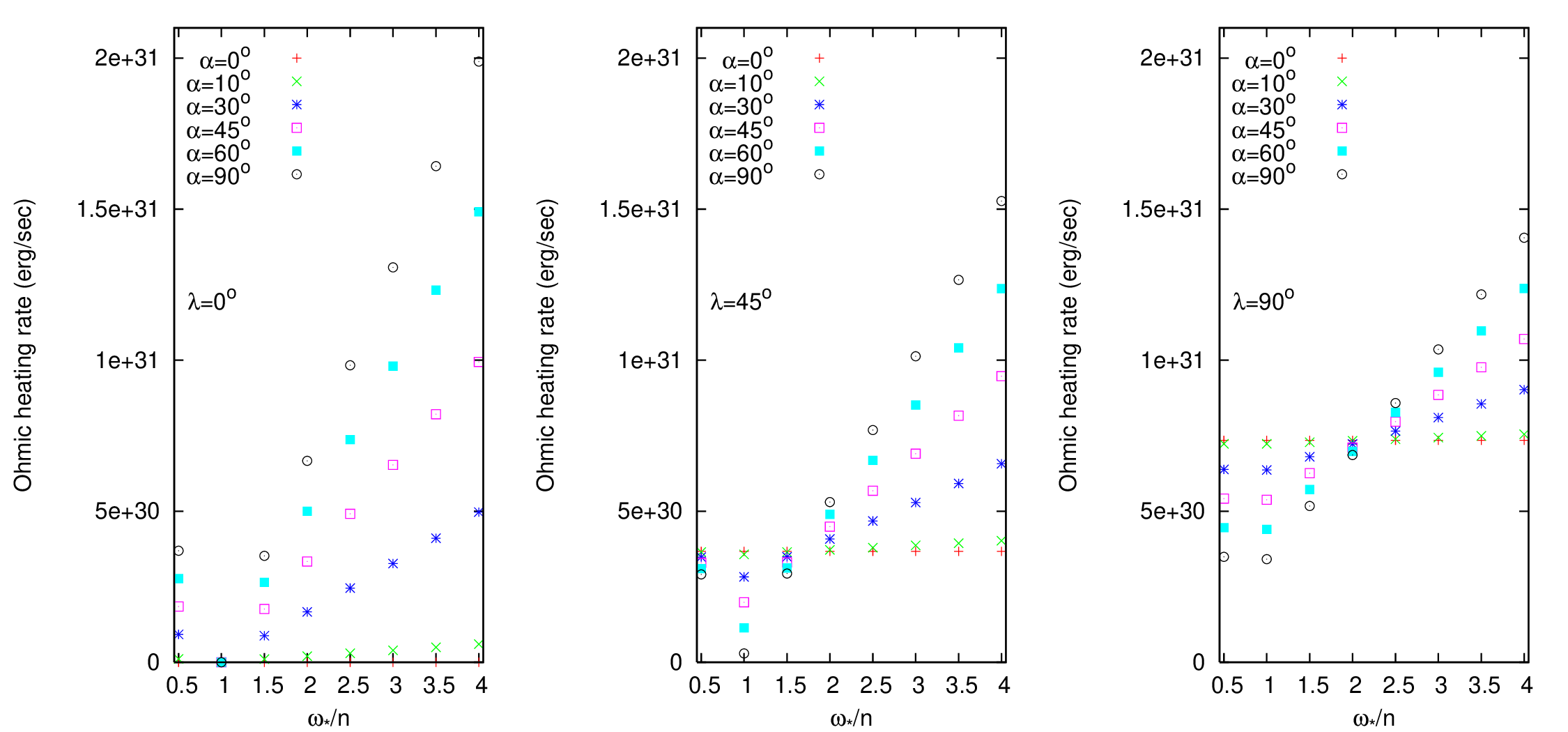}
\includegraphics[scale=.45,angle=-90]{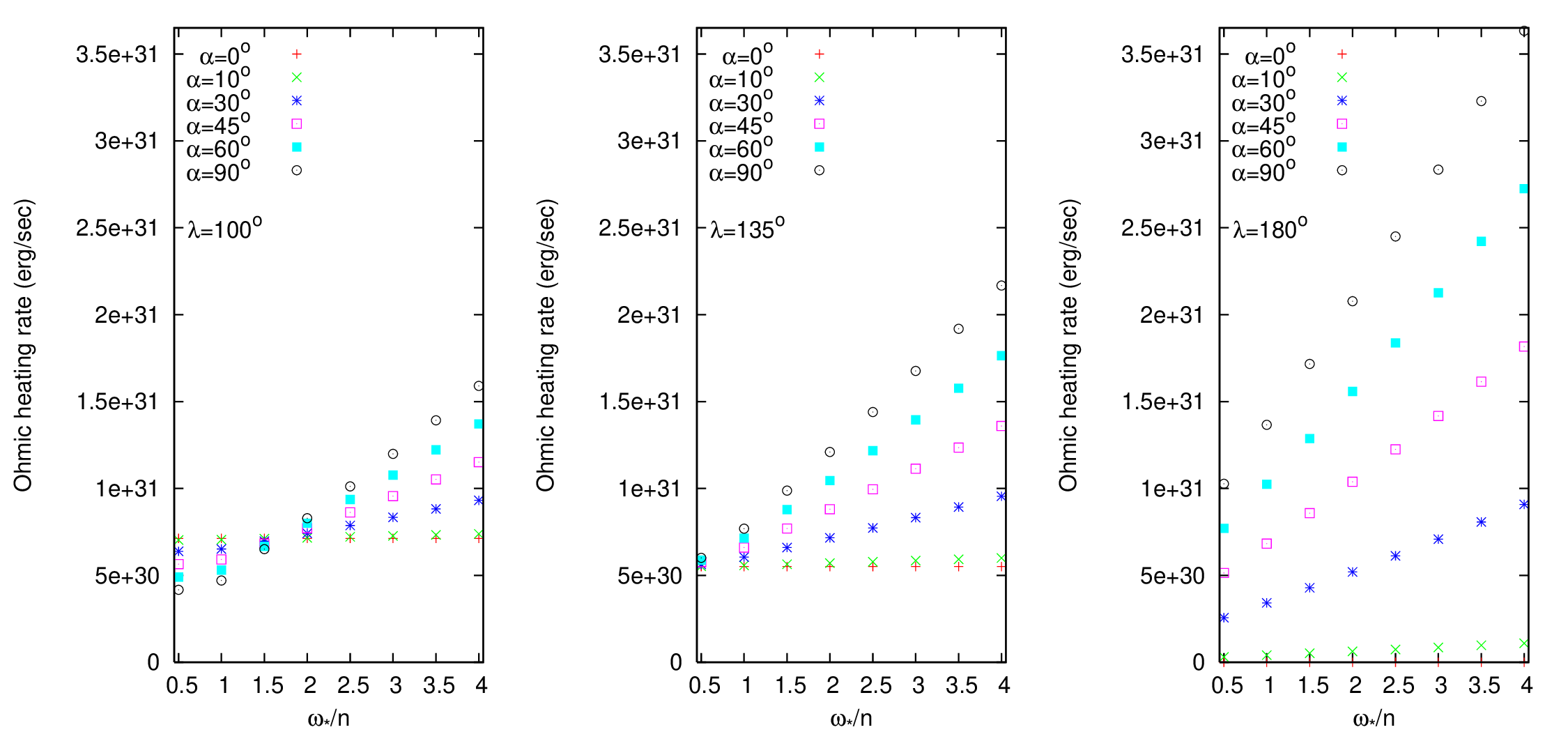}
\caption{Same as Figure \ref{fig2} but the Ohmic heating rates are computed from
Equation~(\ref{eq:heat_torque}).\label{fig4}}
\end{figure}

To further validate our numerical calculations, we also compute $\dot Q_{ohmic}$
based on the general torque-dissipation relation given by
Equation~(\ref{eq:heat_torque}) and show the results in Figure~\ref{fig4}. In
general, Figure~\ref{fig2} and Figure~\ref{fig4} are consistent with each other.
The discrepancy between the two types of calculations of $\dot Q_{ohmic}$ is $<
7$\%. In addition, the results of $\dot Q_{ohmic}$ can be crudely verified by
Equation~(\ref{eq:skin_depth}). Using $\eta=8.3\times 10^9$ m$^2$/s, which is
approximately the maximum value of the $\eta(r)$ profile in the calculations, we
obtain the skin depth $\delta \approx 1.5 \times 10^9$ cm.
The substitution of this skip depth\footnote{In these calculations, the
radiative-convective interface lies at about 1.295$\times 10^{10}$ cm; i.e.
$1.9\times 10^8$ cm below the photosphere. Hence, the main heating region,
characterized by the $\delta$, extends down to the convection zone.} into
Equation~(\ref{eq:skin_depth}) gives the dissipation rate $\dot Q_{ohmic}\approx
2\times 10^{31}$ erg/s, which is on the similar order of the magnitude to those
shown in Figure \ref{fig2}.







\begin{figure}
\includegraphics[scale=.33,angle=-90]{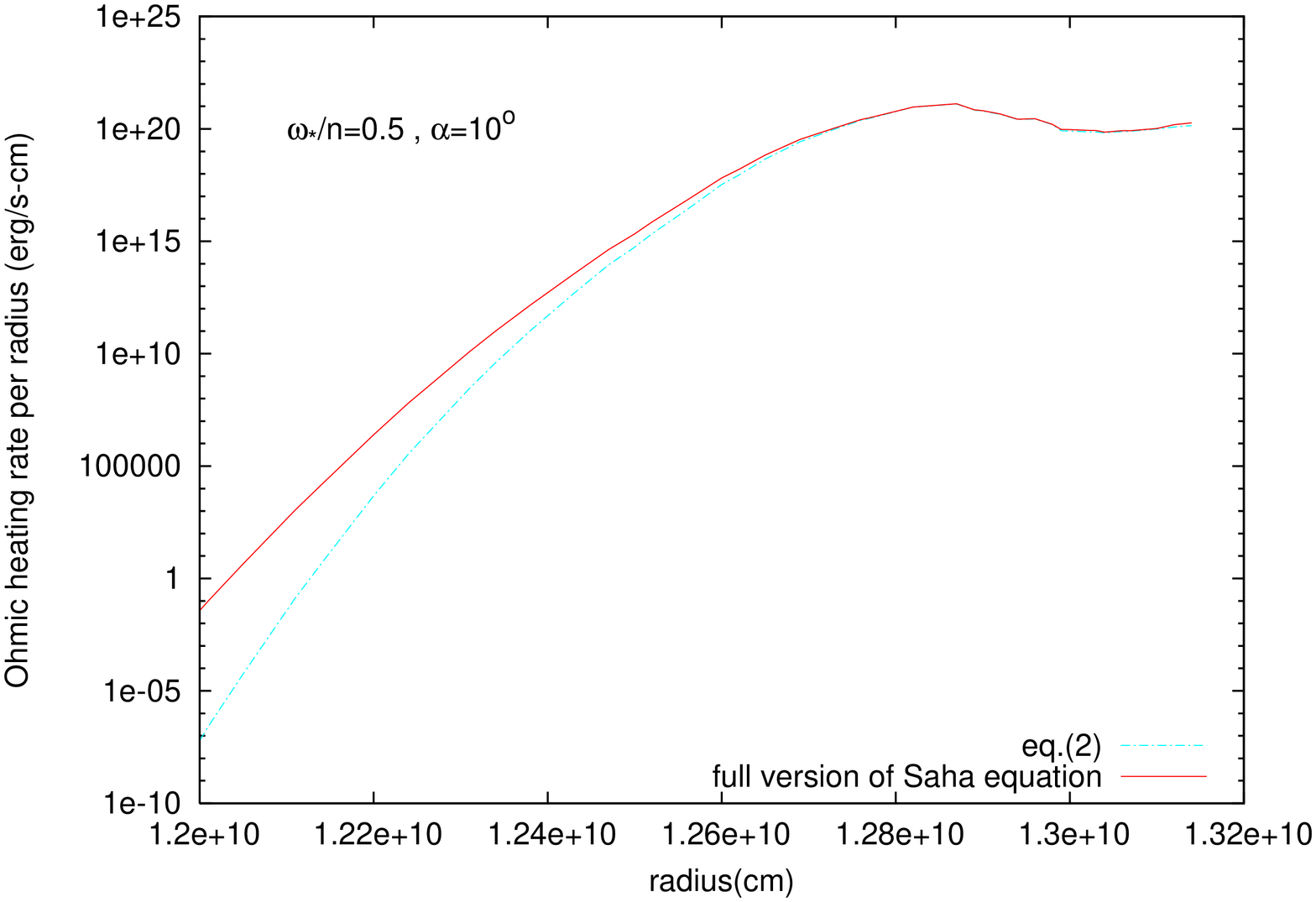}
\includegraphics[scale=.33,angle=-90]{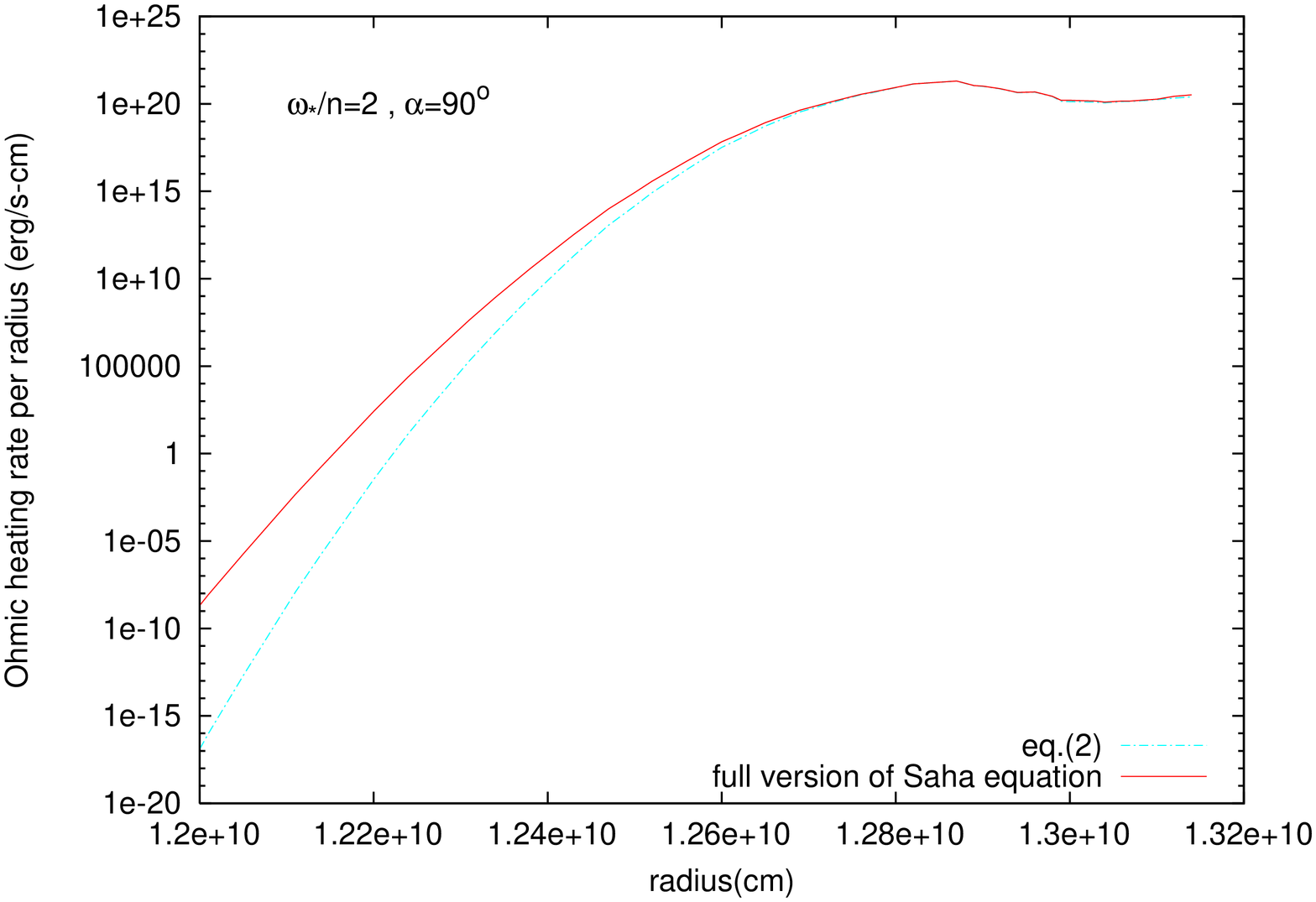}
\caption{Comparison between the total heating profile derived from Equation
(\ref{eq:ionize1}) and that from the full version of the Saha equation for the
cases corresponding to the top left (left in this figure) and bottom right (right
in the figure) panels of Figure \ref{fig5}.\label{fig:compare}}
\end{figure}

All the dissipation rates in the test runs have been calculated based on Equations
(\ref{eq:eta}) and (\ref{eq:ionize1}) under the assumption of the low ionization
fraction of each alkali species as well as the low total ionization fraction within
the skin depth. To verify whether this assumption is reasonable in terms of the
heat generation, we apply the full version of the Saha equation, e.g. Equation(1)
in \citet{BS10}, to the test runs. We find that the total ionization fraction is
sufficiently low in the outer part of the planet such that Equation (\ref{eq:eta})
still applies. We then compute the new heating rate profiles and compare them to
those based on Equation (\ref{eq:ionize1}). Figure \ref{fig:compare} illustrates
the comparisons for the two cases shown in the upper left and lower right panels of
Figure \ref{fig5} as the representative examples. It is evident from the figure
that the heating profiles derived from Equation (\ref{eq:ionize1}) and from the
full version of the Saha equation are almost the same in the outer part of the
planet where most of the dissipation occurs. It then follows that the total heating
rates derived from the full version of the Saha equation are only 3-4\% higher that
those derived from Equation (\ref{eq:ionize1}), thus validating the approximate
results using Equation (\ref{eq:ionize1}).

\section{Evolutionary results}
We now present the evolutionary results with the input parameters and different
initial conditions listed in Table \ref{tbl-1}. The initial $\omega_*$ is given by
one half of the initial $n$. This initial condition is based on the assumptions
that the inner edge of the disk is located at the location of the co-rotation
radius of the proto-star due to disk locking \citep[e.g., see][and reference
therein]{Chang} and that the initial location of the planet lies in the orbit with
the 2:1 mean motion resonance with the inner edge of the disk according to planet
migration theories (see Lin et al. 1996; Rice et al. 2008; cf. Ben\'itez-Llambay et
al. 2011). The simulation is run from $t_i=0.7$ to $t_{end}=10$ Myrs, corresponding
to the T Tauri star phase. The starting time $0.7$ Myrs is comparable to the
timescale of the type II migration time of a giant planet in a protoplanetary disk
\citep{Lin96}. Except for Case 1 which allows $\omega_*$ to evolve according to
Equation~(\ref{eq:ode1}) for comparison, we do not evolve $\omega_*$ in other cases
as it is assumed to be locked by some process such as disk locking. We also run
Case 20 with the parameters similar to the fiducial model in \citet{Laine}:
$M_p=0.63 M_J$, $D_i=0.04$ AU, $\omega_-=10^{-5}$ s$^{-1}$, $\lambda=0^\circ$,
$\alpha=90^\circ$, $M_*=M_\odot$, and $L_*=1.5L_\odot$. It should be stressed that
even in an aligned system, the size of the magnetospheric inner cavity is
proportional to $m^{4/7} {\dot M}^{-2/7}$, where $\dot M$ is the disk gas accretion
rate onto the T Tauri star \citep[e.g. see][and references therein]{Lai99}. In
other words, the initial $n$ is in fact related to $m$ and $\dot M$. Moreover,
$\omega_*$ evolves as the magnetospheric cavity evolves even in the disk-locking
model. Since we assume a constant $m$ and do not intend to model the cavity size in
the presence of stellar obliquity and the misaligned magnetic dipole, we simply
parameterize the initial value of $n$ independent of $m$ in this work.

\begin{figure}
\epsscale{1.1}
\plottwo{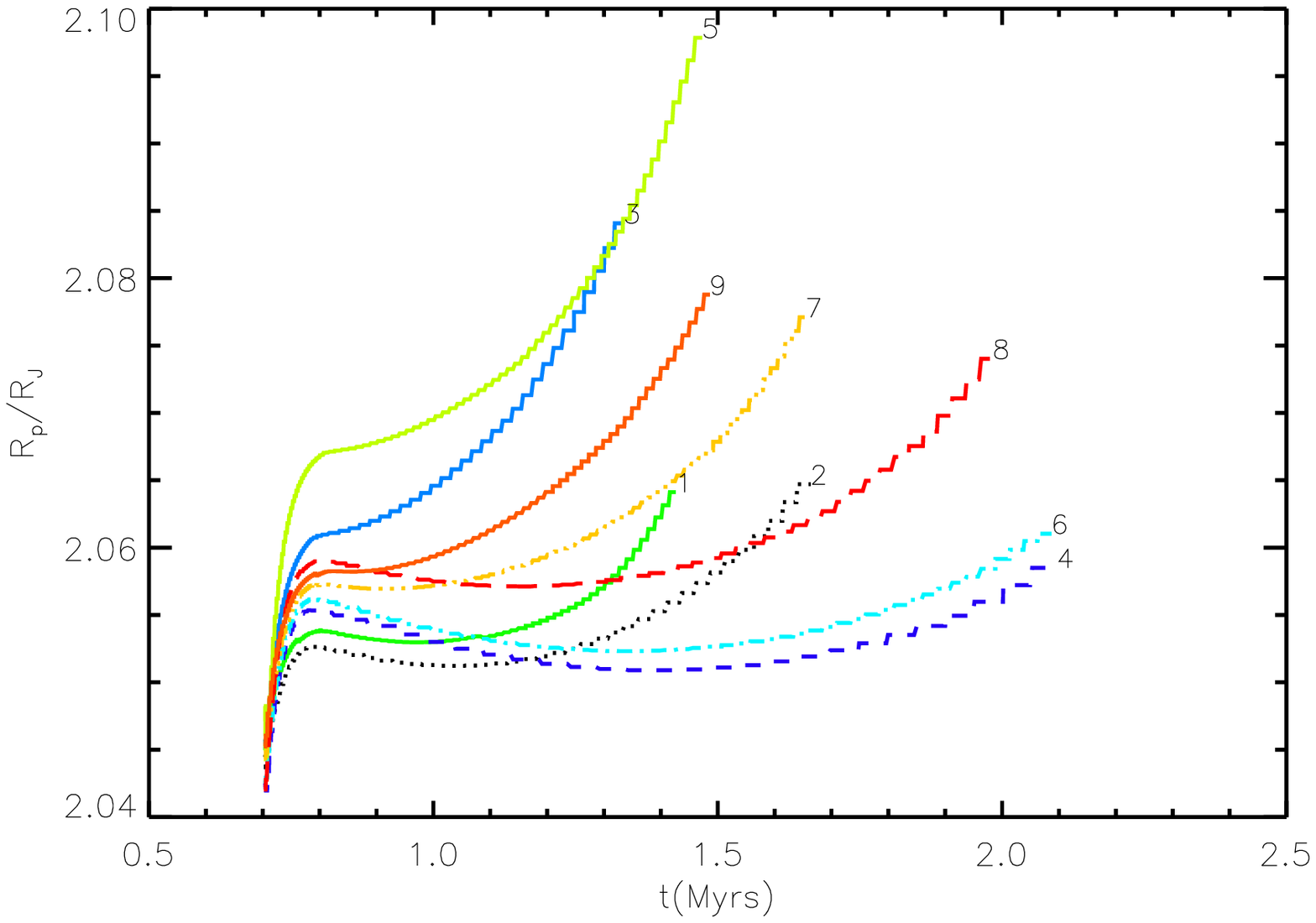}{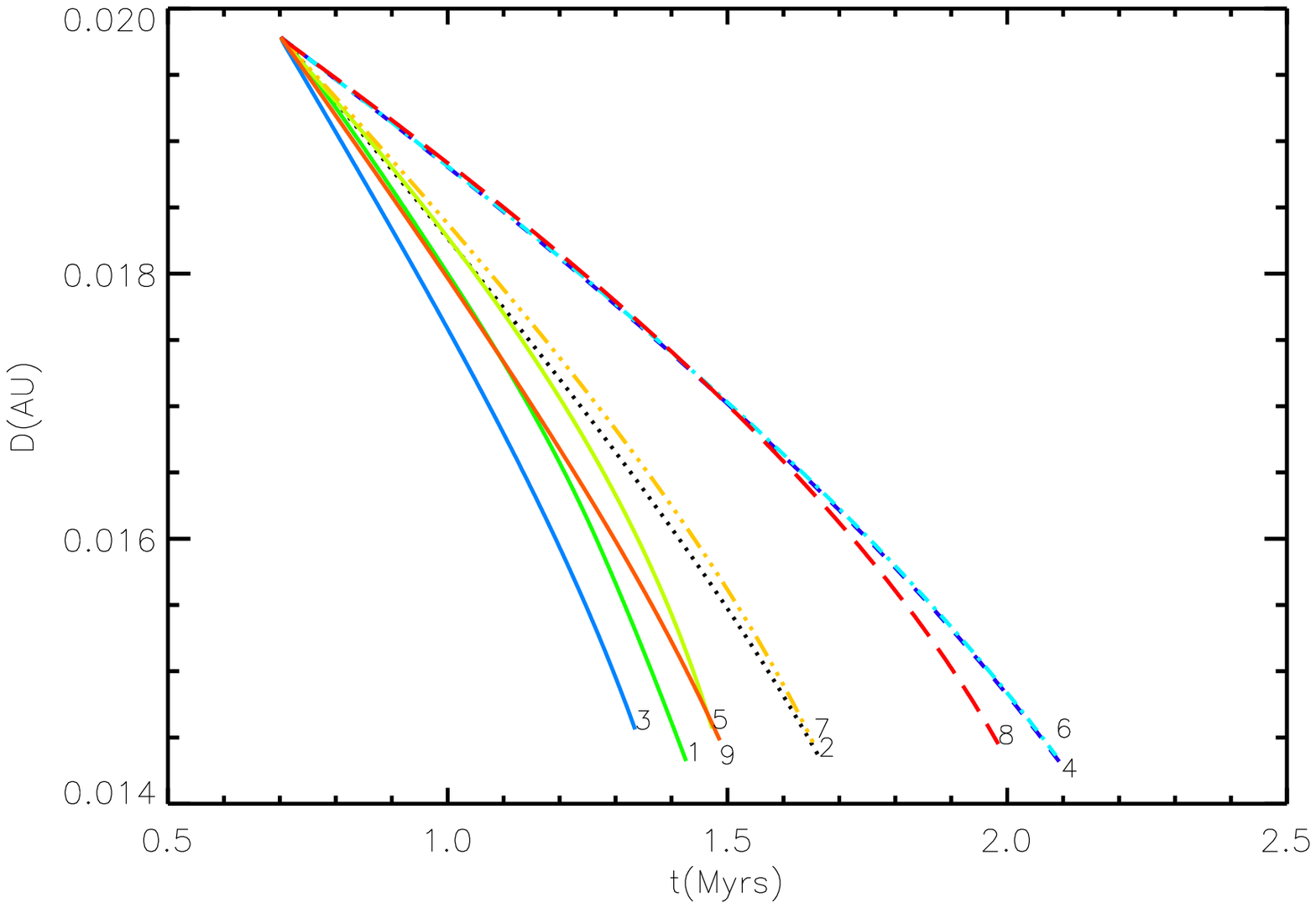}
\caption{Coupled evolutions of $R_p$ and $D$ of the young hot Jupiter in Cases 1-9.
The number labelled next to each curve is the case number. The evolutionary curves
for Cases 4 and 6 are very similar. \label{fig:model1-7a}}
\end{figure}

Cases 1-9 represent the evolutions of a young hot Jupiter of 1 $M_J$ initially at
the very close distance $D_i\approx 0.02$ AU, resembling a planet lying inside a
small magnetospheric cavity.  Owing to the small starting orbital distance, the
strong Ohmic dissipations are generated on the order of $10^{30-31}$ erg/s
throughout the evolutions, resulting in fast orbital decays. However, because the
intense heating occurs mainly near the planet surface, the dissipation is unable to
significantly inflate the planet against self-gravity. Figure~\ref{fig:model1-7a}
shows that the rise in $R_p$ is $<$ 3\% in these cases when the planet quickly
shrinks its orbit and fills its Roche lobe in just a few $10^5$ to about 1 million
years after $t_i$.
The small increase in $R_p$ arises from the thermal expansion of the outer part of
the planet. Despite the intense heating near the planet surface, temperature
inversion is not observed because the strong dissipation is limited to the
radiative layer in the evolutionary cases and thus is easily lost\footnote{It is
different from the non-evolutionary test runs shown in Figures \ref{fig5} and
\ref{fig6} where the heating profiles extend down to the convection zone. When
allowing for evolutions, the Ohmic heating changes the interior structure and thus
decreases $\eta$. As a result of the feedback, the skin depth becomes smaller and
thus the induced electromagnetic effect is mostly confined in the radiative
layer.}. In these cases, the planet first undergoes relatively fast expansion as
the dissipation is suddenly deposited in the beginning, and then reaches an
intermediate quasi-equilibrium state (i.e. $\dot Q_{ohmic} \sim L$ in Equation
\ref{eq:energy}) that lasts for some period of time depending on how fast the orbit
decays. The planet expands again as the orbit continues to shrink and thus the
Ohmic dissipation is further enhanced.

Among these cases, Cases 1, 2, and 3 present the evolutions in the absence of the
stellar obliquity ($\lambda=0$). In Case 1, $\omega_*$ evolves, caused by the
dissipation torques and $\dot I_*$ according to Equation~(\ref{eq:ode1}), without
the spin-locking assumption. In Cases 2 and 3, $\omega_*$ is constant. Figure
\ref{fig:model1-7a} shows that $R_p$ and $D$ in Case 1 evolve faster than those in
Case 2, starting from the same initial conditions. It is because the forcing
frequency $|\omega_-|$ is lower in Case 1, leading to faster orbital decay as
explained in \S3. A larger skin depth results from the slower forcing, generating
deeper heating and thus faster expansion. Figure \ref{fig:model1-7a} also shows
that $R_p$ and $|\dot D|$ in Case 3 are always larger than that in Case 2 as
expected from the test runs in \S3; the larger $\alpha$ in Case 3 produces the
stronger heating and faster orbital decay.

On the other hand, Cases 4, 5, and 6 present the studies in which $\lambda \neq 0$
but the stellar spin and dipole are parallel (i.e. $\alpha=0$). Although the
stellar spins in Cases 4 and 6 point to opposite directions, Figure
\ref{fig:model1-7a} shows that their evolutions of $R_p$ and $D$ are similar due to
the similar time variation of the stellar dipole field that is axi-symmetric about
the spin axis. Furthermore, the larger $R_p$ and $|\dot D|$ in Case 5 than those in
Cases 4 and 6 is a result of the larger $\lambda$ and thus stronger heating, in
accordance with the results of the test runs shown in the upper middle and upper
right panels of Figure~\ref{fig2} for $\alpha=0^\circ$ and $\omega_*/n=0.5$. The
almost symmetric evolutions between Cases 4 and 6 are broken when $\alpha \neq
0^\circ$, as illustrated by the different evolution curves for their counterpart
cases 7 and 8.

We also run Case 9 to compare with Cases 4 and 7 to examine the evolutions starting
from the same $\lambda_i=45^\circ$ but different $\alpha$. As has been demonstrated
in \S3, Case 9 lies in the special regime where $\alpha=90^\circ$ and hence no
dissipation is contributed from the forcing frequency $n$, in contrast to the other
extreme regime shown in Case 4 where the dissipation is totally from the forcing
frequency $n$. The total dissipation rate and therefore $R_p$ as well as $|\dot D|$
increases with $\alpha$ in these cases during the evolutions, which is consistent
with the result of the test run for $\lambda=45^\circ$ and $\omega_*/n=0.5$
displayed in Figure~\ref{fig2}.

Table~\ref{tbl-1} shows that the stellar obliquity $\lambda$ remains zero in Cases
1-3 as expected, because $\langle T_{z'} \rangle$ is the only component of the
dissipation torque acting on the stellar spin. Moreover, $\lambda$ in Cases 4 and 6
change similarly; $\Delta \lambda \approx -11^\circ$ for both cases, meaning that
the dissipation torques turn the system toward the spin-orbit alignment in Case 4
and toward the anti-alignment in Case 6 at similar rates. By contrast, $\lambda$
hardly alters in Case 5 when $\lambda_i=90^\circ$. In the presence of $\alpha$, the
evolution of $\lambda$ is more complicated; the dissipation torques can either
excite or damp the stellar obliquity. Table~\ref{tbl-1} shows that $\lambda$ in
Cases 7 and 8 is decreased by about $2^\circ$-$3^\circ$, while $\lambda$ in Case 9
is increased by about $3^\circ$. The changes of $\lambda$ are nonetheless much
slower than those in Cases 4 and 6.

The aforementioned evolutions of $\lambda$ for Cases 4-9 can be understood through
examination of Equation~(\ref{eq:T_lambda1}). Because $\alpha=0$ in Cases 4-6,
$dL_{spin}/dt=0$ and $dL_{orb}/dt <0$ in these cases\footnote{$dL_{spin}/dt=0$ here
should be distinguished from the assumption $\omega_*=$ constant. $dL_{spin}/dt=0$
in Equation~(\ref{eq:T_lambda1}) results from {\it internal} torques in the
diamagnetic interaction between the star and planet for the cases of $\alpha=0$. On
the other hand, the assumption that $\omega_*=$ constant is made under the
consideration of any {\it external} angular momentum transfer between the T Tauri
star and the environment, such as via disk locking.} as has been shown in \S2.
Equation~(\ref{eq:T_lambda1}) then indicates that the dissipation torques acting on
the star are unable to spin up/down the star but contribute entirely to the
evolution of $\lambda$. This explains why the evolutions of $\lambda$ in Cases 4
and 6 are faster than those in Cases 7, 8, and 9. Besides, we find that the orbital
axis moves faster toward the stellar spin instead of the other way around. It is
due to the fact that $M_p D^2 n/I_*\omega_* \sim 0.1$-$0.01 \ll 1$ in our model.
Namely, it is the terms associated with $1/(M_pD^2n)$ rather than with $1/( I_*
\omega_*)$ on the right-hand side of Equation~(\ref{eq:T_lambda1}) (or equivalently
Equation~(\ref{eq:T_lambda})) that dominate $\dot \lambda$. As a result,
Equation~(\ref{eq:T_lambda1}) gives similar decreases in $\lambda$ in Cases 4 and
6. On the other hand in Case 5, $\lambda_i=90^\circ$ and hence only the terms with
$1/(I_* \omega_*)$ in Equation~(\ref{eq:T_lambda1}) exist at the beginning, which
is negative and small. This explains why $\lambda$ hardly evolves in Case 5;
namely, $\Delta \lambda$ is only $-0.02^\circ$. We can set
Equation~(\ref{eq:T_lambda1}) equal to zero in the case of $\alpha=0$ and find the
equilibrium orientation for the stellar spin, which gives $\lambda = \arccos (-M_p
D^2 n/I_* \omega_*) \approx \pm 90^\circ$, consistent with the small $\dot \lambda$
in Case 5. The equilibrium is unstable as we can anticipate from Cases 4 and 6.
Roughly speaking, the dissipation torque turns the stellar spin and the orbital
axis toward alignment when $|\lambda| \leq 90^\circ$ and toward anti-alignment when
$> 90^\circ$.

When $\alpha \neq 0$, $dL_{spin}/dt$ becomes non-zero. In the case of $n>\omega_*$,
$dL_{spin}/dt >0$ and $dL_{orb}/dt <0$. Equation~(\ref{eq:T_lambda1}) implies that
for Cases 7, 8, and 9, the term associated with $dL_{orb}/dt$ damps $\lambda$,
while the term with $dL_{spin}/dt$ excites $\lambda$. When $\alpha=0$,
$dL_{spin}/dt=0$ and therefore the excitation term vanishes. As $\alpha$ increases
from zero, the excitation term increases as well. This reiterates the point
described in the above paragraph that $\lambda$ damps much more slowly when
$\alpha=45^\circ$. When $\alpha=90^\circ$, it turns out that the excitation term
dominates over the damping term, leading to $\dot \lambda >0$ in Case 9.

As we have seen, the values of $| \lambda |$ in Cases 4 and 6 change similarly.
Indeed, inspection of Figure~\ref{fig:model1-7a} reveals that the evolutions of
$R_p$ and $D$ in Cases 4 and 6 are very similar. As has been described above, the
reason is that their secular interactions are almost the same when the stellar
dipole is aligned or anti-aligned with the stellar spin (i.e. $\alpha=0$).

\begin{figure}
\epsscale{1.1} \plottwo{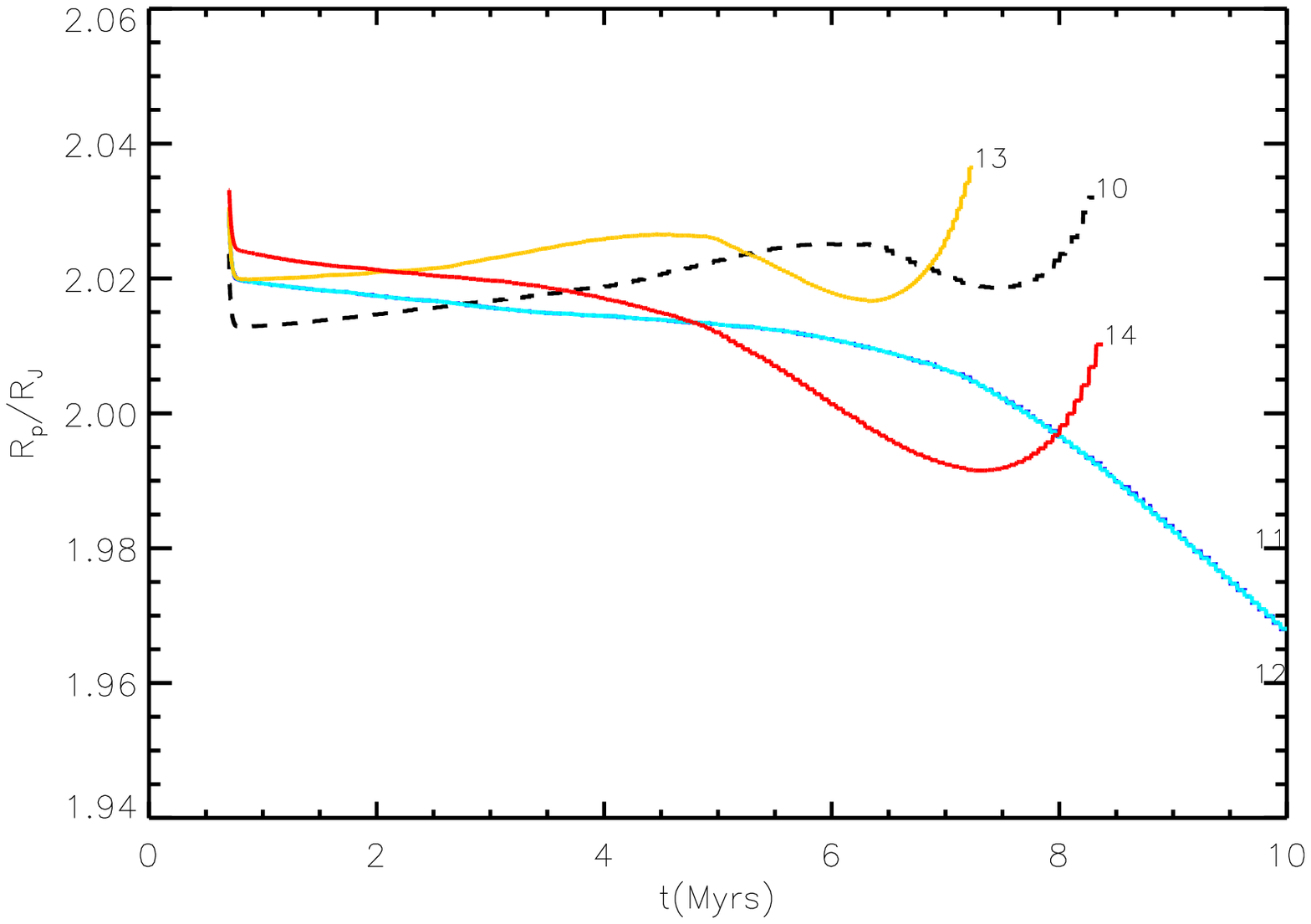}{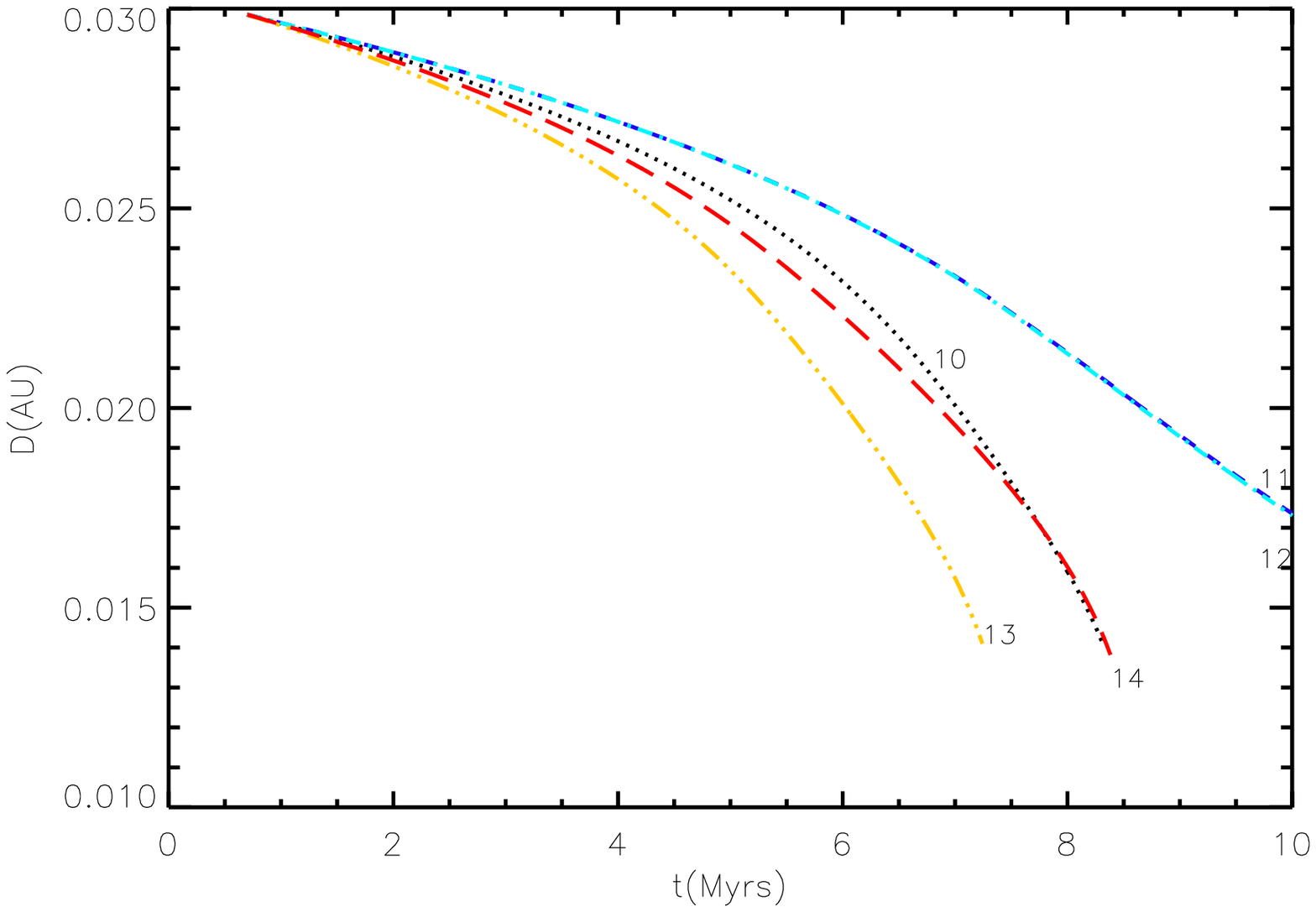}
\caption{Coupled evolutions of $R_p$ and $D$ of the young hot Jupiter in Cases
10-14. The numbers labelled next to the curves indicate the different cases. The
evolutionary curves for Cases 11 and 12 almost overlap.\label{fig:model8-12}}
\end{figure}

Cases 10-14 present the evolutions of the planet placed initially at the farther
distance $D_i\approx 0.03$ AU from its T Tauri star. Hence, the magnetic
interaction and the resulting Ohmic dissipation are weaker at the beginning of the
evolution (i.e. $\dot Q_{ohmic} \sim 10^{29}$ erg/s) than those in Cases 1-9.
Consequently, the planet's orbit decays to $D\lesssim 0.018$ AU over a timescale of
a few Myrs as shown in Figure~\ref{fig:model8-12}. Unlike the initial rise of $R_p$
in Cases 1-9, $R_p$ first decreases quickly over a short period of time, $\sim
10^5$ years in these cases. Then the planet undergoes a slow change of the size
over a few Myrs, meaning that $\dot Q_{ohmic} \sim L$ in Equation (\ref{eq:energy})
such that the radius is more or less maintained by the Ohmic dissipation at $R_p
\sim 2$-$2.04R_J$. In addition, Figure~\ref{fig:model8-12} and Table~\ref{tbl-1}
show that $D$, $R_p$, and $\dot \lambda$ of Cases 11 \& 12 evolve almost
identically, as expected for the same reason as the similarity between Cases 4 \&
6. The stellar obliquities in Cases 11 \& 12 decrease by a similar amount, about
20$^\circ$ at the end of the simulation, but decrease only by about $5^\circ$ in
Cases 13 \& 14. The results are consistent with the comparative study for Cases 4,
6, 7, and 8; i.e., $\lambda$ decays faster when $\alpha=0$ in these cases. The
faster decreases of $\lambda$ in Cases 11 \& 12 give rise to the smaller heating
rates than those in other cases during the late stages of the evolution, leading to
faster contraction and slower orbital decay after $t\approx$ 6-7 Myrs. As a result,
the planet in Cases 11 and 12 is unable to reach Roche-lobe overflow at the end of
the simulation, whereas the planet in Cases 10, 13, and 14 migrates in fast enough
to finally fill the Roche lobe during the T Tauri phase.

\begin{figure}
\epsscale{1.1} \plottwo{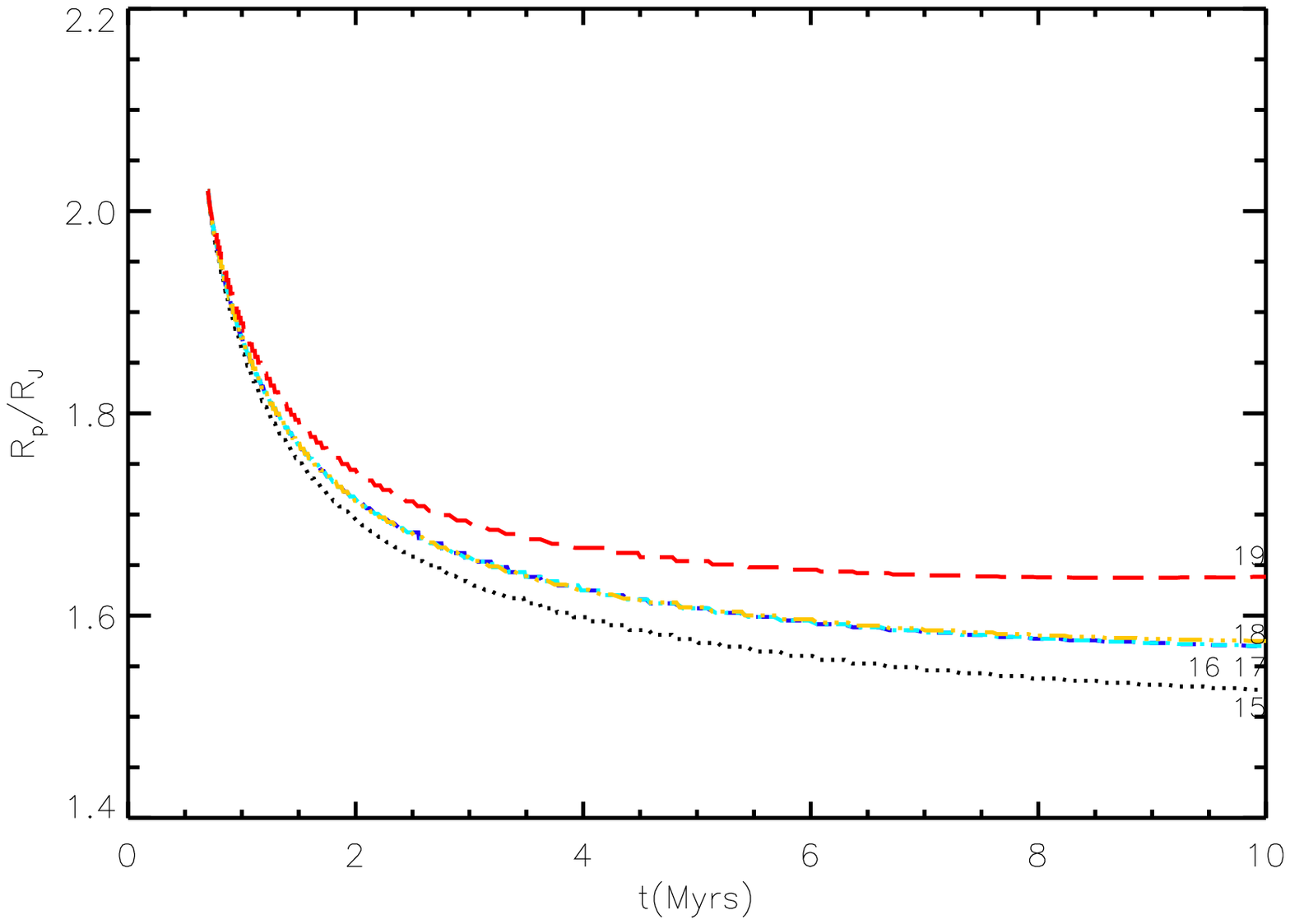}{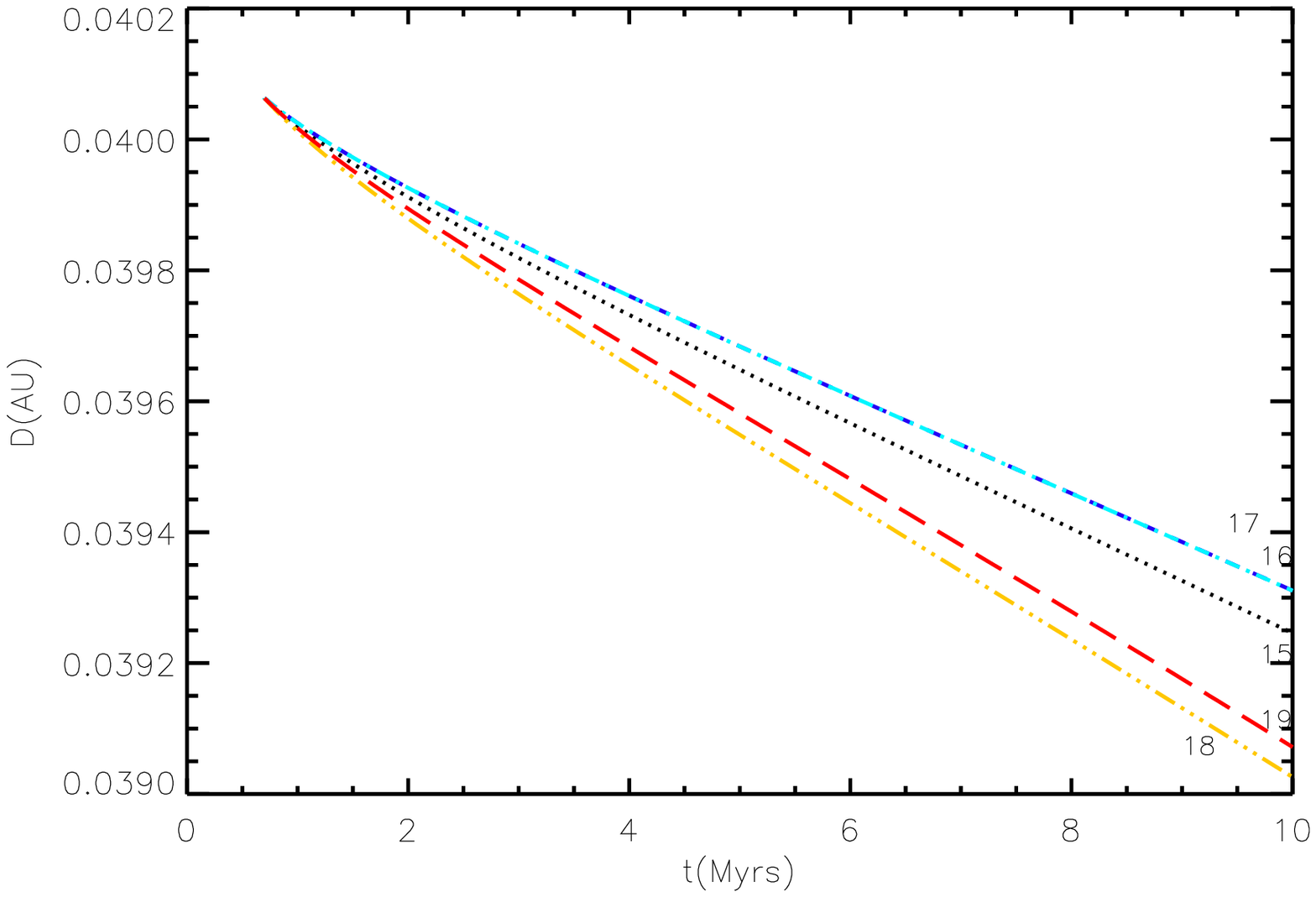}
\caption{Coupled evolutions of $R_p$ and $D$ of the young hot Jupiter in Cases
15-19.
\label{fig:model13-17}}
\end{figure}

In Cases 15-19, we consider a young hot Jupiter initially on the even larger orbit
at $D_i\approx 0.04$ AU. These cases correspond to the scenario that the planet
lies in a large magnetospheric cavity. The larger the orbital distance is, the
weaker the stellar magnetic fields and the interactions are. In all cases, the
Ohmic dissipation rates are around $10^{28}$ erg/s throughout the simulation. As a
result, the young planet first contracts and then gradually attains quasi-thermal
equilibrium at small radii $R_p \approx 1.6 R_J$ after $t\sim 4$-5 Myrs, as
illustrated in Figure~\ref{fig:model13-17}. The orbits do not decay to the distance
$D<0.039$ AU from the star. Hence, the planet never reaches its Roche radius at the
end of the simulation. The corresponding changes in $\lambda$ shown in
Table~\ref{tbl-1} are much smaller as well in these cases.

\begin{figure}
\epsscale{1.1} \plottwo{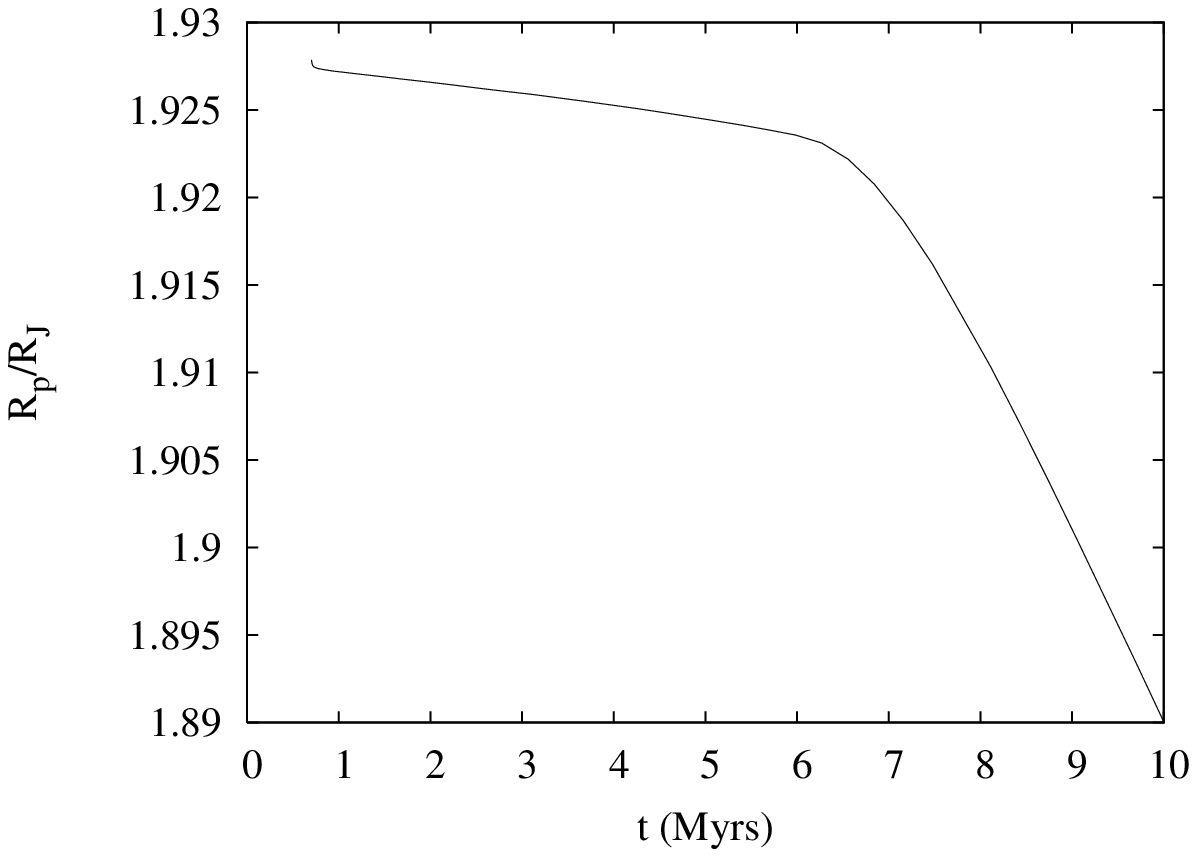}{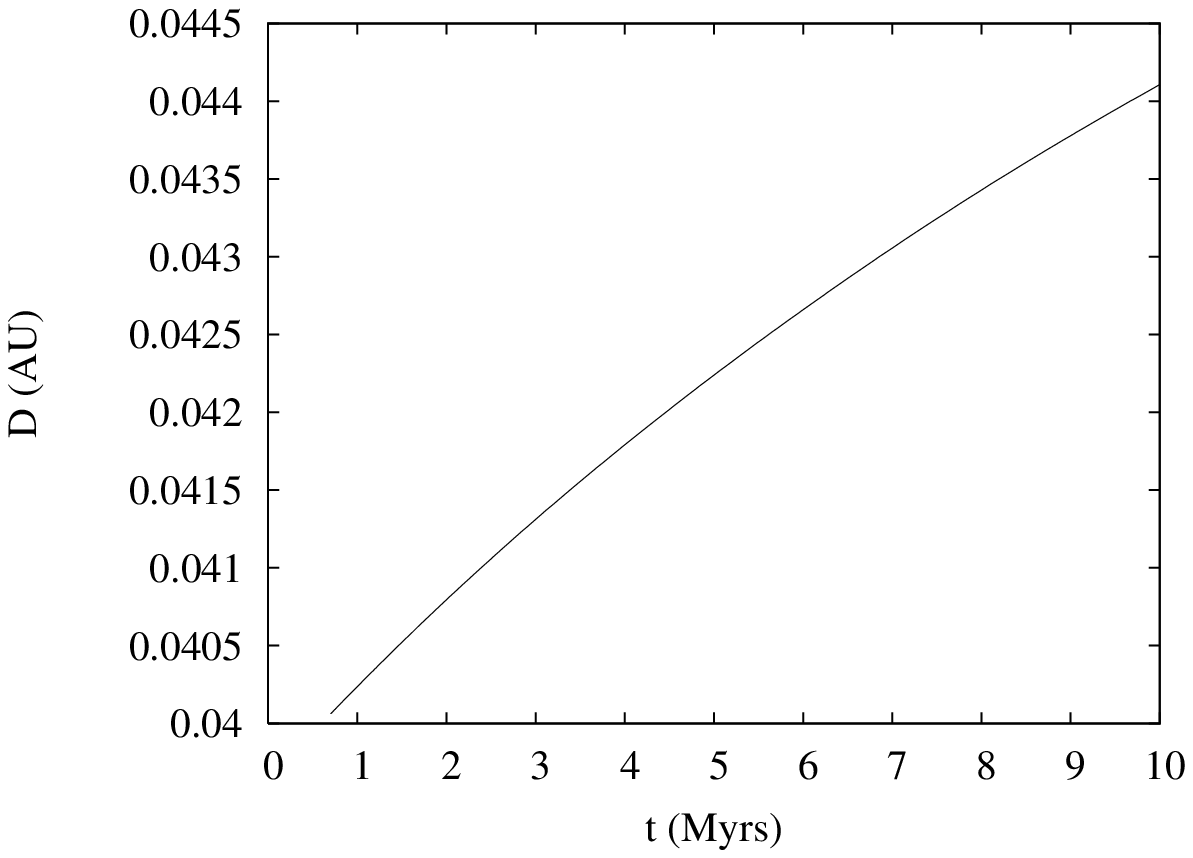}
\caption{Coupled evolutions of $R_p$ and $D$ of the young hot Jupiter in Case 20.
\label{fig:model18}}
\end{figure}

\citet{Laine} focused on a planet that already fills its Roche lobe at $D_i=0.04$
AU (i.e. $R_p\approx 5R_J$), and calculated the resulting dissipation rate and mass
loss rate. The interior structure of the planet is composed of an isothermal
envelope and a polytropic core in their study, which does not truly take into
account the energy equation. Although they suggested that the dissipation can
inflate the planet and trigger mass loss through Roche lobe overflow, the authors
cautioned that the initial condition of the Roche-lobe filling planet would not be
valid. In Case 20, we apply our interior structure to the fiducial model of
\citet{Laine} with the initial $R_p=2.045R_J$ at $t=t_i$. The evolutions of $R_p$
and $D$ are plotted in Figure~\ref{fig:model18}. Since the planet is located
outside the co-rotation orbital radius, the dissipation torque induced by the
forcing $\omega_-
>0$ drives the planet to migrate outwards. Although the planet in this case is less
massive ($M_p=0.63M_J$) than in other cases ($M_p=1M_J$), the Ohmic heating is not
strong enough to inflate the less massive planet to its Roche radius as suggested
by \citet{Laine}. Rather, the planet radius remains almost constant throughout the
T Tauri phase.

\section{Summary and discussions}
We revisit the magnetic interaction between a hot Jupiter and its T Tauri star
investigated by \citet{Laine}. In the original work, the authors considered a
Roche-lobe sized hot Jupiter without its own magnetic fields. The stellar spin was
assumed to be aligned with the orbital axis; i.e. the stellar obliquity
$\lambda=0$. As the planet orbits its parent star, the Ohmic dissipation in the
planet is induced by the stellar magnetic dipole tilted away from the stellar spin
with an angle $\alpha$. To calculate the electric resistivity of the planet, the
interior structure was modelled as a sphere consisting of a polytropic core and an
isothermal outer layer. In their model, the planet lies outside the co-rotating
orbital radius such that the forcing frequency $\omega_->0$. Based on their
fiducial model, the authors suggested that the dissipation torques are not able to
cause any significant orbital change. Nevertheless, the Ohmic heating occurring in
the outer part of the planet is intense enough to inflate the planet up to the
Roche radius. The mass loss through the Lagrangian 1 point toward the central star
provides the angular momentum to the planet and thus possibly halts the planet
migration in the disk.

Motivated by a wide range of the stellar obliquity detected in hot-Jupiter systems
\citep[e.g.][]{Winn11}, which in theory could be excited during the T Tauri phase
\citep{Lai11,FL11}, we extend the original model by considering the coupled
evolution of the interior structure, planet's orbit, and the stellar spin in the
presence of the stellar obliquity $\lambda$. We focus on the secular evolution due
to the dissipation torques and show that the Ohmic dissipation in the planet can be
contributed linearly from the forcing associated with 4 frequencies: $\omega_+$,
$\omega_*$, $n$, and $\omega_-$. Owing to the complication of the problem involving
multiple frequencies, we begin with a couple of test runs based on a given interior
structure for the dissipation calculation, which are further validated by the
general torque-dissipation relation given by Equation~(\ref{eq:heat_torque}) as
well as the skin-depth estimation using Equation~(\ref{eq:skin_depth}).

The coupled evolutions are then carried out for a number of cases listed in
Table~\ref{tbl-1} for the purpose of parameter studies. The evolutions are computed
from $t=0.7$ to $10$ Myrs. The radius of the coreless hot Jupiter of $1M_J$ is
$2.045R_J$ at the beginning. Initially, the T Tauri star is assumed to spin at a
rate $\omega_*=n/2$ to imitate the final stage of the planet migration scenario
with a giant planet inside the magnetospheric cavity of the disk (Lin et al. 1996;
Rice et al. 2008; cf. Ben\'itez-Llambay et al 2011). Since the planet lies inside
the co-rotating orbit, the planet continues to migrate inwards due to the Ohmic
dissipation in the planet. Without modelling the star-disk magnetic interactions,
the co-rotating orbit is simply assumed to correspond to the inner edge of the
magnetically truncated disk despite the presence of $\lambda$ and $\alpha$. This is
certainly one of the limitations of the study. In most of the cases, $\omega_*$ is
assumed constant to simply resemble any processes, such as disk locking, that
maintain the stellar spin.

Three initial orbital distances $D_i \approx 0.02$, 0.03, and 0.04 AU are
considered. With our input parameters for $D_i\approx 0.02$ AU, the intense
dissipation confined near the planet surface only enlarges $R_p$ by $< 3$\%.
Nonetheless, the dissipation torques decay the orbit to the Roche zone in just a
few $10^5$ to about 1 million years. The torques also evolve $\lambda$ when
$\lambda_i \neq 0$. Since $M_pD^2 n \ll I_* \omega_*$, $\dot \lambda$ is primarily
contributed from the movement of the orbital axis rather than the stellar spin
axis. When $\alpha$ is zero, there exists an unstable equilibrium for the
orientation of the stellar spin axis, which points roughly about $90^\circ$ from
the orbital axis. Consequently, the dissipation torques direct the orbital axis
toward the stellar spin for a prograde orbit but toward the anti-parallel direction
to the spin for a retrograde orbit. When $\alpha$ is non-zero, the orbital axis and
the stellar spin can either evolve toward alignment/anti-alignment for small
$\alpha$ or become more misaligned for $\alpha \lesssim 90^\circ$. Because the
stellar spin is not spun up/down by the dissipation torque when $\alpha=0$, it
follows that the stellar obliquity evolves more quickly when the stellar spin is
parallel/anti-parallel to the stellar magnetic dipole.

For the young hot Jupiter initially at the farther distance $D_i \approx $ 0.03 AU,
the dissipation is modest but still strong enough to more or less sustain the
initial $R_p$ except for the cases with $\alpha=0$. The relatively fast decrease of
$\lambda$ for $\alpha=0$ weakens the dissipation and the resulting torques, leading
to the planet contraction and slow orbital decay in the late stage of the
evolution. Therefore, in terms of the cases we have studied, the planet with
$\alpha \neq 0$ undergoes substantial orbital decay in a few Myrs and finally
overflows the Roche-lobe, while the planet with $\alpha=0$ can shrink its orbit but
not sufficiently to allow for Roche-lobe overflow.

Owing to the weaker interaction at the larger orbital distance, the planet of 1
$M_J$ in all the cases starting from $D_i \approx 0.04$ AU contracts and then
roughly reaches quasi-thermal equilibrium during the T Tauri phase, with the final
$R_p$ smaller than those in the cases for $D_i\approx$ 0.02 and 0.03 AU. The
corresponding orbital decays and the changes in the stellar obliquity are
substantially smaller. The planet moves barely from its initial orbit and thus is
not able to reach its Roche-lobe. We also carry out the simulation for the fiducial
model in \citet{Laine} and find that the Ohmic heating can only sustain the radius
of the less massive young hot Jupiter ($M_p=0.63 M_J$), rather than thermally
expanding the planet to its Roche radius as suggested by the authors.


The induced dissipation rates are as high as $10^{30-31}$ erg/s when the planet
moves to about $D <$ 0.02 AU. The intense heating near the planet's surface does
not generate local temperature inversion in our model, in contrast to the surface
heating models presented in \citet{Gu04} and \citet{WL12}. It is probably because
the dissipation responsible for the temperature inversion lies within a thin shell
fairly deep in Gu et al. 2004 (a prescribed narrow Gaussian region) and in Wu \&
Lithwick 2012 (a region at about the optical depth of 100), while the dissipation
in our diamagnetic induction model is deposited so close to the surface that it is
easily lost and so there is no local maximum in T.

In this work, we introduce the planet at $t_i=0.7$ Myrs and adopt a constant
magnetic dipole moment.
In theory, a gas giant planet can form later and migrate to the magnetospheric
cavity at a later time \citep[e.g.,][]{il09,Mordasini}. Besides, the magnetic
dipole moment may decay over the course of a few Myrs \citep{JK,YJ11}. Hence, our
results probably give the suggestive values for the maximum changes of the stellar
obliquity, orbital distance, and planet radius during the T Tauri star phase.

The orbital decay of a young hot Jupiter in our magnetic model is not significant
unless the distance to the T Tauri star is smaller than about 0.03 AU. Some other
process, such as gravitational tides \citep{Chang} or a small magnetospheric cavity
during FU-Orionis outbursts \citep{Baraffe09,Adams09}, can bring the planet in to
that distance. The Ohmic mechanism could be the final stage in bringing a planet in
very close to the T Tauri star or even leading to a Roche-lobe overflow. This may
provide one of the explanations for the pile-up of hot Jupiters with the orbital
periods of $\sim$ 3 days, and could also reduce the too-high population of hot
Jupiters inferred from the population synthesis model \citep{il09}. In this work,
we do not study the post-evolution of a Roche-lobe filled planet. In terms of our
model parameters, a young hot Jupiter located within 0.03 AU from its T Tauri star
can undergo fast orbital decay on the timescales much shorter than 10 million
years. Therefore it is possible that the
 young planet overflows its Roche lobe, migrates out, then migrates in, and overflows again. Consequently, the
planet suffers from intermittent mass losses until its density is low enough to go
through the stage of the runaway adiabatic mass loss \citep{Chang}, leading to the
demise of the planet during the T Tauri phase.

As has been described in the Introduction, there is a large body of literature
devoted to a variety of magnetic interactions between a hot Jupiter and its parent
star, some of which can also cause the angular momentum to transfer between the
planet's orbit and the stellar spin. The efficiency of angular momentum transfer is
model dependent, relying on the Ohmic dissipation rate. In the T Tauri phase, the
presence of a disk is expected to magnetically affect the stellar spin and perhaps
the planet's orbit. \citet{Lai11} and \citet{FL11} considered a hybrid magnetic
model including diamagnetic induction and magnetic-field linkage for the purpose of
the generation of the stellar obliquity. In the study presented here, we follow the
work by \citet{Laine} and therefore focus only on the diamagnetic interaction
between the planet and its T Tauri star. Our simple model suggests that the stellar
obliquity starting from a non-zero value may further evolve after the planet
migrates into the magnetospheric cavity, making the orbit of the young hot Jupiter
incline with the disk plane. As a result, a hot Jupiter does not necessarily lie on
the same orbital plane with the planets farther out from the central star. Whether
or not our model can provide a wide range of stellar obliquities at the end of T
Tauri phase depends on the initial distribution of stellar obliquity as well as the
distribution of the direction of stellar dipole moment relative to the stellar
spin.

In the Introduction, we also caution that the skin depth beneath the photosphere is
one of the major uncertainties of the model. In the presence of a planetary
ionosphere or magnetosphere, the value may be appreciably smaller than what we
compute in this work. Nevertheless, as an analog of the star-disk magnetic
interactions \citep{Lai99}, our diamagnetic model and other magnetic interactions
should be considered together for the orbital evolution inside the magnetospheric
cavity. While theoretical models are under development, it is conceivable in the
future that photometric variability on timescales of a few days
\citep[e.g.][]{Bouvier} and spectropolarimetry applied to T Tauri stars
\citep[e.g.][]{Donati,Long} would serve as possible detection methods, to search
for the variability modes and magnetic perturbations that are associated with the
orbital motion of such a young hot Jupiter during the T Tauri star stage.

\acknowledgments We thank Shi-Shin Chang for providing us with the evolution of
$I_*$ of a T Tauri star. We benefit from the discussions with Gordon I. Ogilvie
about the relation between the dissipation and torques in the presence of the
stellar obliquity. Y. C. and P. G. were supported by an NSC grant in Taiwan through
NSC 100-2112-M-001-005-MY3. P. B. was
supported by an NSF grant AST0908807.

\appendix
\section{magnetic scalar potential in the presence of stellar obliquity}

We work on the problem with two sets of the coordinate systems $O'x'y'z'$ and
$Oxyz$ illustrated in Figure \ref{fig1}, which were employed in \citet{Laine} with
the orbital angular velocity and stellar spin around the $z$ (and $z'$ axis). The
magnetic dipole moment $\bf m$ is tilted with the angle $\alpha$ from the stellar
spin and thus can be expressed as ${\bf m}=(\mu_0/4\pi)(B_*R_*^3/2) {\bf \hat m}$
with the unit dipole moment vector ${\bf \hat m}=\sin \alpha \cos \omega_*t {\bf
i'} + \sin \alpha \sin \omega_* t {\bf j'} + \cos \alpha {\bf k'}$ as viewed in the
star's frame. Here $B_*$ is the surface stellar field and $R_*$ is the stellar
radius. Then let the stellar spin rotate along the $x'$ axis clockwise
(counter-clockwise) on the $y'$-$z'$ plane such that the stellar obliquity angle is
$\lambda >0$ ($<0$). This gives \mbf{\omega_*} in the inertial frame of our
problem. Note that the sign definition of $\lambda$ agrees with that used for the
Rossiter-McLaughlin effect.

We then rotate the stellar spin along the $z'$ axis clockwise at the angular
velocity $n$ to give the stellar spin as observed in the co-moving frame of the
planet. As a result, the unit vector of the stellar magnetic dipole moment $\bf
\hat m_{obliquity}$ as viewed in the planet's rest frame becomes
\begin{equation}
 {\bf \hat m_{obliquity}} =  {\bf P_{z'}P_{x'} \hat m} =
 \begin{pmatrix}
\sin \alpha (\cos \omega_* t \cos nt + \cos \lambda \sin \omega_* t \sin nt) + \sin \lambda \cos \alpha \sin nt \\
\sin \alpha (-\cos \omega_* t \sin nt + \cos \lambda \sin \omega_* t \cos nt)+ \sin \lambda \cos \alpha \cos nt \\
-\sin \lambda \sin \alpha \sin \omega_* t + \cos \lambda \cos \alpha
\end{pmatrix}\,
\label{eq:m_obliquity}
\end{equation}
The rotation matrices involved in the above equations are
\[
{\bf P_{x'}} = \left( \begin{array}{ccc}
1 & 0 & 0 \\
0 & \cos \lambda & \sin \lambda \\
0 & -\sin \lambda & \cos \lambda \end{array} \right), \qquad {\bf P_{z'}} = \left(
\begin{array}{ccc}
\cos nt & \sin nt & 0 \\
-\sin nt & \cos nt & 0 \\
0 & 0 & 1 \end{array} \right).
\]


Note that when the obliquity $\lambda=0$ in Equation~(\ref{eq:m_obliquity}), we
recover ${\bf \hat m}=\sin \alpha \cos \omega t {\bf i'} + \sin \alpha \sin \omega
t {\bf j'} + \cos \alpha {\bf k}$ with the Doppler-shifted frequency $\omega$ being
$\omega_*-n$ as viewed in the co-moving frame of the planet. Moreover, if the
obliquity is retained but the magnetic axis is aligned with the stellar spin (i.e.
$\alpha=0$), we obtain ${\bf \hat m_{obliquity}}=\sin \lambda \sin n t {\bf i'} +
\sin \lambda \cos n t {\bf j'} + \cos \lambda {\bf k}$ as should be expected in the
co-moving frame of the planet.

Then we can express the magnetic scalar potential due to the stellar magnetic
dipole moment with non-zero stellar obliquity in the co-moving frame of the plane
as follows \citep[cf.][]{Campbell83}:
\begin{eqnarray}
V_*&=&{\mu_0 m \over 4\pi r'^3} {\bf r'} \cdot {\bf \hat m}_{obliquity} \nonumber\\
&=&{\mu_0 m \over 4\pi r'^3} [ r\sin \theta \sin \varphi \sin \alpha (\cos \omega_*
t \cos nt+ \cos \lambda \sin nt \sin \omega_* t)+
r\sin \theta \sin \varphi \sin nt \sin \lambda \cos \alpha \nonumber\\
&&+ (D-r\sin \theta \cos \varphi )\sin \alpha (-\sin nt \cos \omega_*
t + \cos \lambda \cos nt \sin \omega_* t) \nonumber\\
&&+ (D-r\sin \theta \cos \varphi) \sin \lambda \cos \alpha \cos nt + r\cos \theta
(-\sin \lambda \sin
\alpha \sin \omega_* t + \cos \lambda \cos \alpha )] \nonumber \\
&\approx &{\mu_0 m r\over 4\pi D^3}\{-P_1^1[{1\over 2}\sin \varphi \sin \alpha
((\cos \omega_+ t +\cos \omega_- t)+ \cos \lambda (\cos \omega_- t-\cos \omega_+
t)) \nonumber\\
&&+\sin \varphi \sin nt \sin \lambda \cos \alpha \nonumber\\
&&+ \cos \varphi \sin \alpha ((\sin \omega_- t -\sin \omega_+ t) + \cos \lambda
(\sin
\omega_+t + \sin \omega_- t) ) \nonumber\\
&&+ 2\cos \varphi
\sin \lambda \cos \alpha \cos nt] -P_1^0 \sin \lambda \sin \alpha \sin \omega_* t \}\nonumber\\
&&+{3\mu_0 m r^2 \over 8\pi D^4}\{P_2^2 [ {1\over 6} \sin 2\varphi \sin \alpha (
(\cos \omega_+t + \cos \omega_-t) + \cos \lambda
(\cos \omega_- t - \cos \omega_+ t)) \nonumber\\
&&+{1\over 3} \sin 2\varphi \sin nt \sin \lambda \cos \alpha \nonumber\\
&&+ {1\over 4} \cos 2\varphi \sin \alpha ((\sin \omega_-t-\sin \omega_+t) +\cos
\lambda (\sin \omega_+t+\sin \omega_- t)) \nonumber\\
&& + {1\over
2} \cos 2\varphi \sin \lambda \cos \alpha \cos nt]\nonumber \\
 &&- P_2^0 [ {1\over 2}\sin \alpha ((\sin \omega_- t -\sin \omega_+ t) + \cos \lambda
 (\sin \omega_+ t + \sin \omega_- t)) + \sin \lambda \cos \alpha \cos nt] \nonumber\\
&&+{2\over 3} P_2^1 \cos \varphi \sin \lambda \sin \alpha \sin \omega_* t \}
\end{eqnarray}
where $\omega_+\equiv \omega_* +n$, $\omega_-\equiv \omega_* -n$, and the
approximation of $V_*$ is obtained by the expansion of $r'^{-3}$ up to the order of
$(r/D)^2$. In the above equation, the time-independent terms are dropped out
because they do not contribute to the electromagnetic induction. Besides,
$P_l^{|m|}$ are associated Legendre functions, defined by\footnote{Note that one
can obtain the expression for $m<0$ from $m>0$ using the relation
$P_l^{-m}(x)=(-1)^m {(l-m)! \over (l+m)!} P_l^m(x)$.} $P_1^0=\cos \theta$,
$P_1^1=-\sin \theta$, $P_2^0=(1/2)(3\cos^2 \theta -1)$, $P_2^1=-3\sin\theta
\cos\theta$, and $P_2^2=3\sin^2 \theta$.

Using $V_*=-\partial_r \phi_*$, we have the poloidal scalar of the stellar field as
follows
\begin{eqnarray}
\phi_*&=& {\mu_0 m r^2\over 8\pi D^3}\{P_1^1[{1\over 2}\sin \varphi \sin \alpha
((\cos \omega_+ t +\cos \omega_- t)+ \cos \lambda (\cos \omega_- t-\cos \omega_+
t))\nonumber \\
&&+\sin \varphi \sin nt \sin \lambda \cos \alpha \nonumber\\
&&+ \cos \varphi \sin \alpha ((\sin \omega_- t -\sin \omega_+ t) + \cos \lambda
(\sin
\omega_+t + \sin \omega_- t) )\nonumber \\
&&+ 2\cos \varphi
\sin \lambda \cos \alpha \cos nt] +P_1^0 \sin \lambda \sin \alpha \sin \omega_* t \}\nonumber\\
&&+{\mu_0 m r^3 \over 8\pi D^4}\{ -P_2^2 [ {1\over 6} \sin 2\varphi \sin \alpha (
(\cos \omega_+t + \cos \omega_-t) + \cos \lambda
(\cos \omega_- t - \cos \omega_+ t)) \nonumber \\
&&+{1\over 3} \sin 2\varphi \sin nt \sin \lambda \cos \alpha \nonumber \\
&&+ {1\over 4} \cos 2\varphi \sin \alpha ((\sin \omega_-t-\sin \omega_+t) +\cos
\lambda (\sin \omega_+t+\sin \omega_- t))\nonumber \\
&& + {1\over
2} \cos 2\varphi \sin \lambda \cos \alpha \cos nt]\nonumber \\
 &&+ P_2^0 [ {1\over 2}\sin \alpha ((\sin \omega_- t -\sin \omega_+ t) + \cos \lambda
 (\sin \omega_+ t + \sin \omega_- t)) + \sin \lambda \cos \alpha \cos nt]\nonumber\\
&&-{2\over 3} P_2^1 \cos \varphi \sin \lambda \sin \alpha \sin \omega_* t
\}.\label{eq:phi_*}
\end{eqnarray}
Note that we recover the original form of the poloidal scalar $\phi_*$ in
\citet{Laine} when $\lambda=0$. In addition, because there is a poloidal scalar
outside the planet $\phi_p$ generated by the fields induced by $\phi_*$ inside the
planet, $\phi_p$ has the same time and angular dependence as $\phi_*$. Hence,
$\phi_p$ is given by
\begin{eqnarray}
&&\phi_p(r\geq R_p,t)=
{\mu_0 P_1^0 \over r} (\delta_1 \sin \omega_* t + \delta_2 \cos \omega_* t) \nonumber\\
&&+\mu_0 P_1^1 [ {\sin \varphi \over r} (\alpha_1 \sin \omega_+ t + \alpha_2 \cos
\omega_+ t + \alpha_3 \sin \omega_- + \alpha_4 \cos \omega_-t + \alpha_5 \sin nt +
\alpha_6 \cos nt) \nonumber \\
&& + {\cos \varphi \over r}(\alpha_7 \sin \omega_+ t + \alpha_8 \cos \omega_+ t +
\alpha_9 \sin \omega_- t
+\alpha_{10} \cos \omega_- t + \alpha_{11} \sin nt + \alpha_{12} \cos nt) ] \nonumber\\
&&+{\mu_0 P_2^0 \over r^2} (\beta_1 \sin \omega_+ t + \beta_2 \cos \omega_+ t +
\beta_3 \sin \omega_- t+\beta_4 \cos \omega_- t + \beta_5 \sin nt + \beta_6 \cos nt) \nonumber\\
&&+\mu_0 P_2^1  {\cos \varphi \over r^2} (\epsilon_1 \sin \omega_* t + \epsilon_2
\cos \omega_* t )  \nonumber \\
&&+\mu_0 P_2^2 [ {\sin 2\varphi \over r^2} (\gamma_1 \sin \omega_+ t + \gamma_2
\cos \omega_+ t+ \gamma_3 \sin \omega_- t + \gamma_4 \cos \omega_- t + \gamma_5
\sin nt
+ \gamma_6 \cos nt ) \nonumber \\
&&+ {\cos 2\varphi \over r^2}(\gamma_7 \sin \omega_+ t + \gamma_8 \cos \omega_+ t +
\gamma_9 \sin \omega_- t + \gamma_{10} \cos \omega_- t + \gamma_{11} \sin nt +
\gamma_{12} \cos nt) ].
\end{eqnarray}

The above form for $\phi_*$ along with the separation of variables for the
solutions to the induction equation suggest that
the induced potential scalar $\phi_p$ can be in general given by
\begin{eqnarray}
\phi({\bf r},t)&=&\Sigma_{l,m} C{_l^m}_{,\omega_+} G_{l,\omega_+}(r) Y_l^m(\theta,
\varphi)
 e^{i\omega_+t}+\Sigma_{l,m} C{_l^m}_{,\omega_-} G_{l,\omega_-}(r) Y_l^m(\theta, \varphi) e^{i\omega_- t}\nonumber\\
 &&+\Sigma_{l,m} C{_l^m}_{,\omega_*} G_{l,\omega_*}(r)
 Y_l^me^{i\omega_* t}
 +\Sigma_{l,m} C{_l^m}_{,nt} G_{l,n}(r) Y_l^m(\theta, \varphi)e^{int}.
\end{eqnarray}
When $\lambda=0$, the above expression is reduced to only one term that is
associated with $\omega_-$.

Inside the planet ($r\leq R_p$), the $r$-dependence of $\phi_p$ is given by
\begin{equation}
\left[ {d^2 \over dr^2} -\left( {l(l+1) \over r^2} +{i\omega \over \eta} \right)
\right] G_{l,\omega}(r)=0,
\end{equation}
where the forcing frequency $\omega$ denotes $\omega_+$, $\omega_-$, $\omega_*$, or
$n$. If $\eta(r)$ is known, the above equation with each frequency can be solved by
rearranging the equation to 4 first-order ODEs as have shown in Equation~(10) of
\citet{Laine}, subject to the boundary conditions
\begin{equation}
G'_l(R_p)+{l\over R_p} G_l(R_p)
-(2l+1)R_p^l=0,
\end{equation}
\begin{equation}
G'_l(r\approx 0)-{l+1\over r}
G_l(r\approx 0)=0,
\end{equation}

After obtaining the solutions for $G_{l,\omega}$, we are ready to solve for the
coefficients in the expressions of $\phi_p(r\leq R_p,t)$ and $\phi_p(r\geq R_p,t)$
by demanding the condition that the potential scalars and their derivatives should
continue at $r=R_p$ for each frequency. Namely, $\phi_*+\phi_{p,out}=\phi_{p,in}$
and $\partial_r (\phi_*+\phi_{p,out})=\partial_r \phi_{p,in}$ at $r=R_p$ for each
frequency. As a result, the terms associated with $\omega_+$ contribute 10
algebraic equations from $Y_1^{\pm 1} e^{i\omega_+ t}$, $Y_2^0 e^{i\omega_+ t}$,
and $Y_2^{\pm 2} e^{i\omega_+ t}$. Likewise, the terms depending on $\omega_-$ and
$n$ give rise to 10 algebraic equations each. On the other hand, the terms with
$\omega_*$ only contribute 6 algebraic equations from $Y_1^0 e^{i\omega_* t}$ and
$Y_2^{\pm 1}e^{i\omega_* t}$, but keep in mind that the two terms associated with
$\sin \varphi$ from $Y_2^{\pm 1}e^{i\omega_* t}$ should vanish as they do not exist
in $\phi_p$, leading to the relation between $C{_2^1}_{,\omega_*}$ and
$C{_2^{-1}}_{,\omega_*}$ (i.e. $6C{_2^1}_{,\omega_*}=C{_2^{-1}}_{,\omega_*}$) and
hence reducing to 4 equations. Including their derivatives counterparts, we shall
solve 20 algebraic equations for the 20 coefficients associated with $\omega_+$,
$\omega_-$, and $n$.\footnote{The 20 equations for $\omega_-$ are equivalent to the
set of 20 linear equations in the Appendix B of \citet{Laine}. There are typos on
the right-hand sides of their 5th to 7th linear equations.} Besides, 8 equations
for the 8 coefficients associated with $\omega_*$. In the end, we have 68 algebraic
equations at $r=R_p$ and solve for the 68 coefficients, which are $\delta_{1,2}$,
$\alpha_{1,\cdots 12}$, $\beta_{1,\cdots 6}$, $\epsilon_{1,2}$, $\gamma_{1,\cdots
12}$, Re$(C{_1^{\pm 1}}_{,\omega_+,\omega_-,n})$, Im$(C{_1^{\pm
1}}_{,\omega_+,\omega_-,n})$, Re$(C{_2^{0}}_{,\omega_+,\omega_-,n})$,
Im$(C{_2^{0}}_{,\omega_+,\omega_-,n})$, Re$(C{_2^{\pm 2}}_{,\omega_+,\omega_-,n})$,
Im$(C{_1^{\pm 1}}_{,\omega_+,\omega_-,n})$, Re$(C{_1^{0}}_{,\omega_*})$,
Im$(C{_1^{0}}_{,\omega_*})$, Re$(C{_2^{1}}_{,\omega_*})$, and
Im$(C{_2^{1}}_{,\omega_*})$.
Then we know $\phi_{p,out}$ and $\phi_{p,in}$.


\section{Secular evolutions of the precession and obliquity of the stellar spin}
While the stellar spin and the planet's orbit exchanges angular momentum due to the
dissipative torques in our magnetic model, the total angular momentum vector is
conserved. Bearing this in mind, we have the secular evolution of the precession
angle governed by \citep[cf.][]{GP70}
\begin{equation}
\sin \lambda_* I_* \omega_* {d \langle \varphi'' \rangle \over dt} = \langle
T_{x''} \rangle,\label{eq:precession_A}
\end{equation}
where $\lambda_*$ is the angle between the stellar spin and the total angular
momentum. Since the stellar spin angular momentum, orbital angular momentum, and
total angular momentum form a vector triangle, it is straightforward to show from
the triangle that
\begin{equation}
\sin \lambda_* = {\sin \lambda \over \sqrt{(I_*\omega_*/M_pD^2n)^2+1+
2(I_*\omega_*/M_p D^2 n)\cos \lambda}} \equiv f(\lambda).
\end{equation}
This equation yields the expression of $f(\lambda)$ in
Equation(\ref{eq:precession}).

Now we turn to the evolution of the stellar obliquity. We define $d\lambda_*/dt$
and $d\lambda_n/dt$ as the contributions to the secular evolution of $\lambda$ due
respectively to the stellar spin and orbital angular momentum moving toward/away
from the total angular momentum. As has been described in the main text, $\dot
\lambda$ is caused by the components of $\langle T_{y''} \rangle$ and $\langle
T_{z'} \rangle$ normal to the stellar spin and by the back reaction $-\langle
T_{y''} \rangle$ acting to the orbital angular momentum. Hence we have
\citep[cf.][]{GP70,Lai99}
\begin{equation}
I_* \omega_* {d\lambda_*\over dt}= \langle T_{y''}\rangle \cos \lambda - \langle
T_{z'} \rangle \sin \lambda,
\end{equation}
\begin{equation}
M_p D^2 n {d\lambda_n \over dt} = \langle T_{y''} \rangle.
\end{equation}
Therefore the equation
\begin{equation}
{d\lambda \over dt}={d\lambda_* \over dt} + {d\lambda_n \over dt},
\end{equation}
gives the expression in Equation~(\ref{eq:T_lambda}).

When $M_pD^2 n \gg I_* \omega_*$, $\lambda_* \approx \lambda$ and $\dot \lambda_*
\approx \dot \lambda$ as they ought to be because the total angular momentum is
almost contributed from the orbital angular momentum. However, $M_p D^2 n < I_*
\omega_*$ for a system consisting of a hot Jupiter and a T Tauri star. As a result,
the planet's orbit can evolve more significantly than the stellar spin.

\clearpage

\begin{table}
\caption{model calculations. $m=4\times 10^{34}$ A m$^2$,
$M_p=1 M_J$, $M_*=1M_{\odot}$, $L_*=L_\odot$. The subscript $i$ denotes the initial
value, while the subscript $f$ means the final value due to either the planet
reaching its Roche lobe or $t=t_{end}$. In addition, at $t_i$, $R_p=2.045R_J$,
$\langle \phi'' \rangle=0$. The column ``overflow" indicates in which cases the
planet reaches the Roche-lobe overflow before $t=t_{end}$. The figure number for
the results of each case is indicated in the column ``Figure".\label{tbl-1}}
\begin{tabular}{cccrrcc}
\tableline
 Case & $D_i$ (AU) & $\alpha$ & $\lambda_i$ &$\lambda_f$ & overflow & Figure \\
\tableline\tableline

1 & 0.02 & $45^\circ$ & $0^{\circ}$  & 0$^\circ$ & yes & \ref{fig:model1-7a} \\
2 & 0.02 & $45^\circ$ & $0^{\circ}$ & 0$^\circ$ & yes & \ref{fig:model1-7a}  \\
3 & 0.02 & $90^{\circ}$& $0^{\circ}$ & 0$^\circ$ & yes & \ref{fig:model1-7a}\\
4 & 0.02 & $0^{\circ}$ & $45^{\circ}$ &$33.77^{\circ}$ & yes  &\ref{fig:model1-7a} \\
5 & 0.02 & $0^{\circ}$ & $90^{\circ}$ &$89.98^{\circ}$ & yes &\ref{fig:model1-7a} \\
6 & 0.02 & $0^{\circ}$ & $-135^{\circ}$ & $-146.02^\circ$ & yes &\ref{fig:model1-7a} \\
7 & 0.02 & $45^{\circ}$ & $45^{\circ}$  &43.26$^\circ$ & yes &\ref{fig:model1-7a} \\
8 & 0.02 & $45^{\circ}$ & $-135^{\circ}$  & $-137.97^\circ$ & yes &\ref{fig:model1-7a} \\
9 & 0.02 & $90^{\circ}$ & $45^{\circ}$  & $48.25^\circ$ & yes &\ref{fig:model1-7a} \\
10 & 0.03 & $45^{\circ}$ & $0^{\circ}$ & 0$^\circ$ & yes & \ref{fig:model8-12}\\
11 & 0.03 & $0^{\circ}$ & $45^{\circ}$  & $21.97^\circ$ & no & \ref{fig:model8-12} \\
12 & 0.03 & $0^{\circ}$ & $-135^{\circ}$ & $-157.89^\circ$ & no & \ref{fig:model8-12} \\
13 & 0.03 & $45^{\circ}$ & $45^{\circ}$ & $40.56^\circ$& yes & \ref{fig:model8-12} \\
14 & 0.03 & $45^\circ$ & $-135^\circ$ & $-141.13^\circ$ & yes & \ref{fig:model8-12} \\
15 & 0.04 & $45^{\circ}$ & $0^{\circ}$ & $0^\circ$ & no & \ref{fig:model13-17} \\
16 & 0.04 & $0^{\circ}$ & $45^{\circ}$  & $44.44^\circ$ & no & \ref{fig:model13-17} \\
17 & 0.04 & $0^\circ$ & $-135^\circ$ & $-135.55^\circ$ & no & \ref{fig:model13-17} \\
18 & 0.04 & $45^\circ$ & $45^\circ$ & $44.88^\circ$ & no & \ref{fig:model13-17} \\
19 & 0.04 & $45^\circ$ & $-135^\circ$ & $-135.16^\circ$ & no & \ref{fig:model13-17} \\
20\tablenotemark{a} & 0.04 & $90^\circ$ & $0^\circ$ & $0^\circ$ & no & \ref{fig:model18} \\
\end{tabular}
\tablenotetext{a}{Fiducial model in \citet{Laine}. See the text for the details.}
\end{table}

\clearpage

\clearpage

\clearpage

\clearpage

\clearpage

\clearpage
\clearpage


\clearpage

\clearpage
\end{document}